%% file: main_cooling.tex
\documentclass[12pt,a4paper,leqno]{article}
\usepackage{amsfonts,amsmath,amssymb,amsthm,enumerate,CJK,array}
\usepackage{graphicx}
\usepackage{setspace}
\usepackage{fullpage}
\usepackage[font=small,labelfont=bf,labelsep=colon]{caption}
\usepackage{natbib}
\usepackage{subfig}
\usepackage{booktabs}
\usepackage[para,flushleft]{threeparttable}
\usepackage{adjustbox}
\usepackage{multirow}
\usepackage{lscape}
\usepackage{color}
\usepackage[hidelinks]{hyperref}
\usepackage{siunitx}
\def\E{\mbox{E}}

\newcommand{\vek}[1]{\mathbf{#1}}
\newcommand{\bsy}[1]{\boldsymbol{#1}}

\begin{document}

\title{{COOLING MEASURES AND HOUSING WEALTH: EVIDENCE FROM SINGAPORE}\setlength{\baselineskip}{1.25em}}
\author{Wolfgang Karl H\"{a}rdle \and Rainer Schulz \and Taojun Xie\thanks{H\"{a}rdle: Ladislaus von Bortkiewicz Chair of Statistics, C.A.S.E. -- Center for Applied Statistics, Humboldt-Universit\"{a}t zu Berlin, Germany, and Sim Kee Boon Institute for Financial Economics, Singapore; Schulz: University of Aberdeen Business School, United Kingdom; Xie (corresponding author): Asia Competitiveness Institute, Lee Kuan Yew School of Public Policy, National University of Singapore, x.tj@icloud.com. Financial support of the European Union’s Horizon 2020 research and innovation program “FINTECH: A Financial supervision and Technology compliance training programme” under the grant agreement No 825215 (Topic: ICT-35-2018, Type of action: CSA), the European Cooperation in Science \& Technology COST Action grant CA19130 - Fintech and Artificial Intelligence in Finance - Towards a transparent financial industry, the Deutsche Forschungsgemeinschaft’s IRTG 1792 grant, the Yushan Scholar Program of Taiwan and the Czech Science Foundation’s grant no. 19-28231X / CAS: XDA 23020303 are greatly acknowledged.\setlength{\baselineskip}{1.25em}}}
\maketitle

\begin{abstract}
\noindent Excessive house price growth was at the heart of the financial crisis in 2007/08. Since then, many countries have added \emph{cooling measures} to their regulatory frameworks. It has been found that these measures can indeed control price growth, but no one has examined whether this has adverse consequences for the  housing wealth distribution. We  examine this for Singapore, which started in 2009 to  target price growth over ten rounds in total. We find that welfare from housing wealth in the last round might not be higher than before 2009. This depends on the deflator used to convert nominal into real prices.  Irrespective of the deflator, we can reject that welfare increased monotonically over the different rounds.  
\\[1em]
{\bf JEL Classification}:  R31, C31, C55 \setlength{\baselineskip}{1.25em}
\setlength{\baselineskip}{1.25em}

\end{abstract}

\newpage

\input{cooling_introduction}

\input{cooling_context}

\input{cooling_methodology}
\input{cooling_data}

\input{cooling_results}
\input{cooling_conclusion}

\setcounter{section}{0}
\renewcommand{\thesection}{\Alph{section}}
\renewcommand{\theequation}{\Alph{section}\arabic{equation}}
\setcounter{equation}{0} 

\input{cooling_appendix}

\section*{Acknowledgements}
We thank J\"{u}rgen Bracht, Mauro Papi, Tsur Somerville, Joe Swierzbinski, Verity Watson,
and Axel Werwatz for helpful comments. The usual disclaimer applies. Taojun Xie thanks the Sim Kee Boom Institute for Financial Economics, Lee Kong Chian School of Business, Singapore Management University, for financial support.

\newpage

\bibliographystyle{kluwer}
\bibliography{singapore}

\newpage

\section*{Tables and Figures}

\begin{landscape}
\input{./tables/table_context_descriptives.tex}
\end{landscape}

\input{./tables/table_n_obs.tex}

\begin{landscape}
\input{./tables/descriptive_type.tex}
\end{landscape}

\renewcommand{\thesection}{\Alph{section}}
\setcounter{table}{0}
\renewcommand{\thetable}{\Alph{section}\arabic{table}}

\begin{landscape}
\input{./tables/SD_hypothesis1.tex}
\end{landscape}

\newpage
%

\begin{figure}[htp!]
\centering
\includegraphics[width = \textwidth]{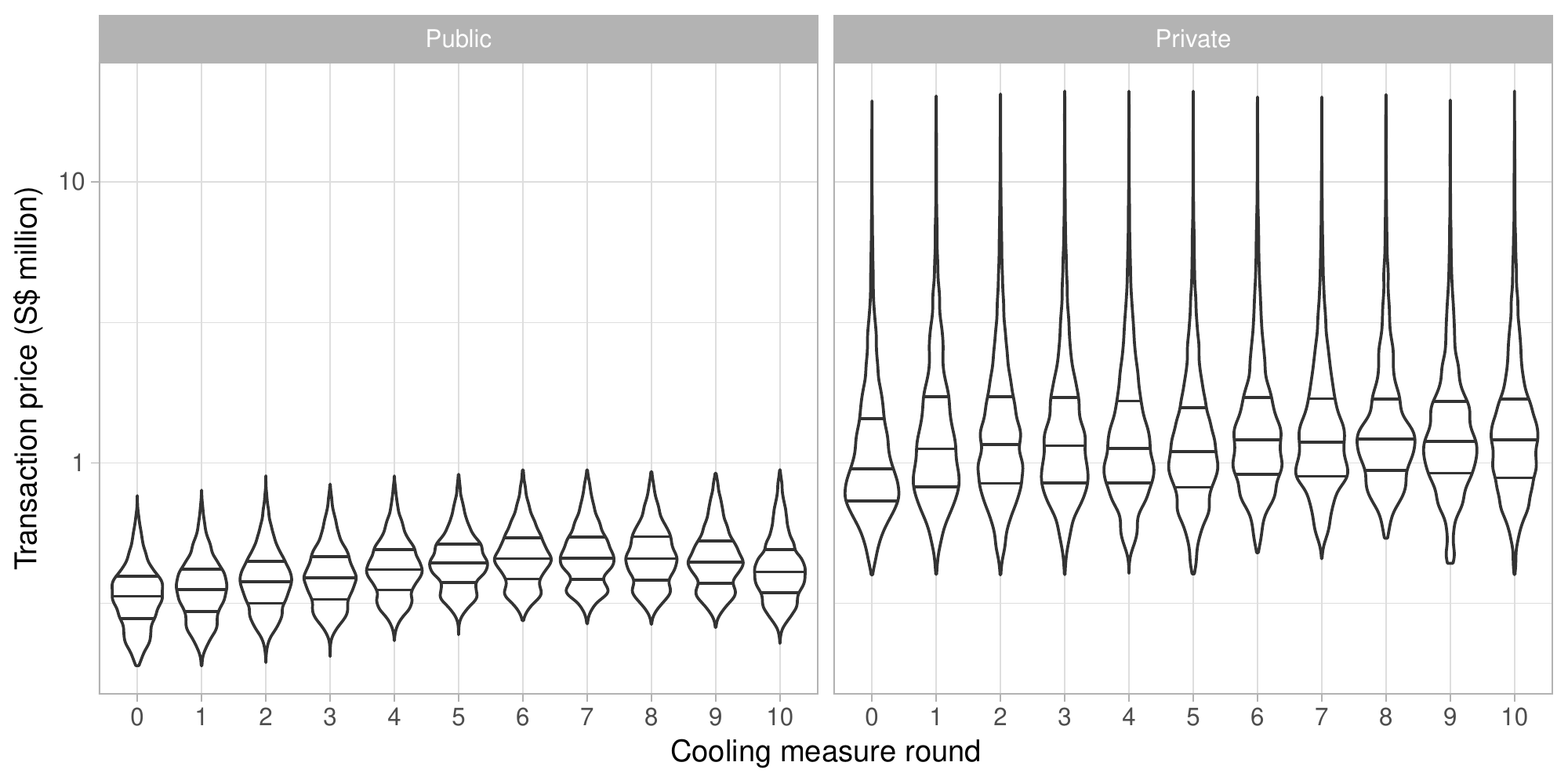}
\caption{{\bf Distribution of nominal transaction prices by cooling measure round.} Shows  violin plots for prices in the public (left) and the private (right) sector for the different cooling measure rounds. Horizontal lines are interquartile range and median, symmetric vertical lines are estimated kernel densities. Round 0 corresponds to the base period.}
\label{fig:violin_nominal_prices}
\end{figure}

\newpage

\begin{figure}[htp!]
\centering
\includegraphics[width = \textwidth]{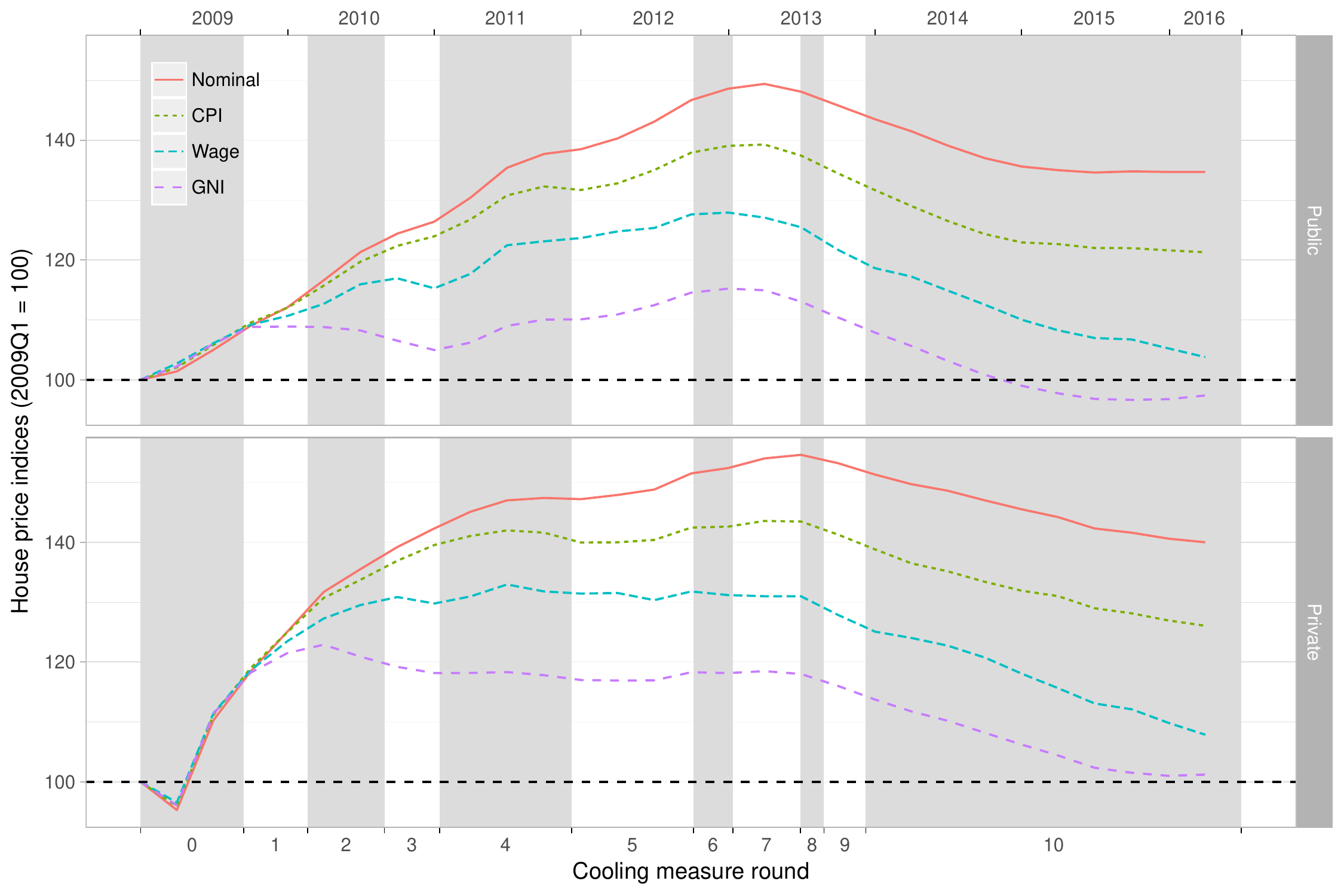}
\caption{{\bf Nominal and real house price trends and cooling measures rounds, Singapore 2009Q1-2016Q2.} Shows the official quarterly transaction price indices for the public (HDB) and the private sector (URA) in nominal terms and deflated with each of the three deflator series. Series are interpolated with cubic splines. The duration of the base period 0 and the ten cooling measure rounds are indicated by the background shading.}
\label{fig:average_prices}
\end{figure}

\begin{figure}
\centering
  \centering
\includegraphics[width = \textwidth]{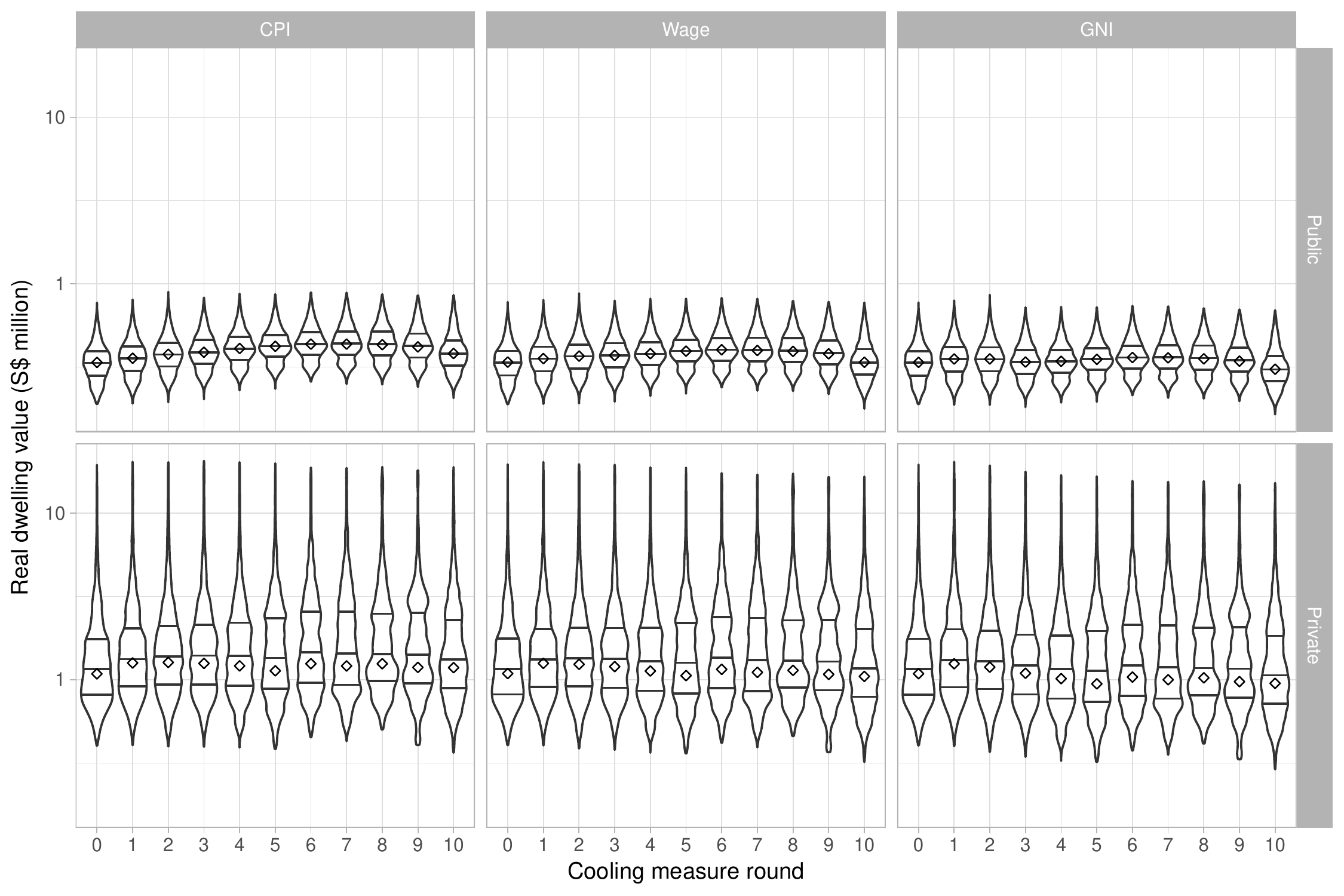}
 \caption{{\bf Distribution of dwelling values by cooling measure round.} Shows  violin plots for the dwelling value distributions in the public  and the private sector for the different cooling measure rounds and the three deflators. The dwelling value distribution re-weights transaction prices to take the composition of the housing stock into account. Horizontal lines in the violin plots are interquartile range and median, symmetric vertical lines are estimated kernel densities. The diamond indicates the arithmetic average. Round 0 corresponds  to the base period.}
\label{fig:violin_real_prices}
\end{figure}

\newpage

\begin{figure}[htp!]
\centering
\includegraphics[width = \textwidth]{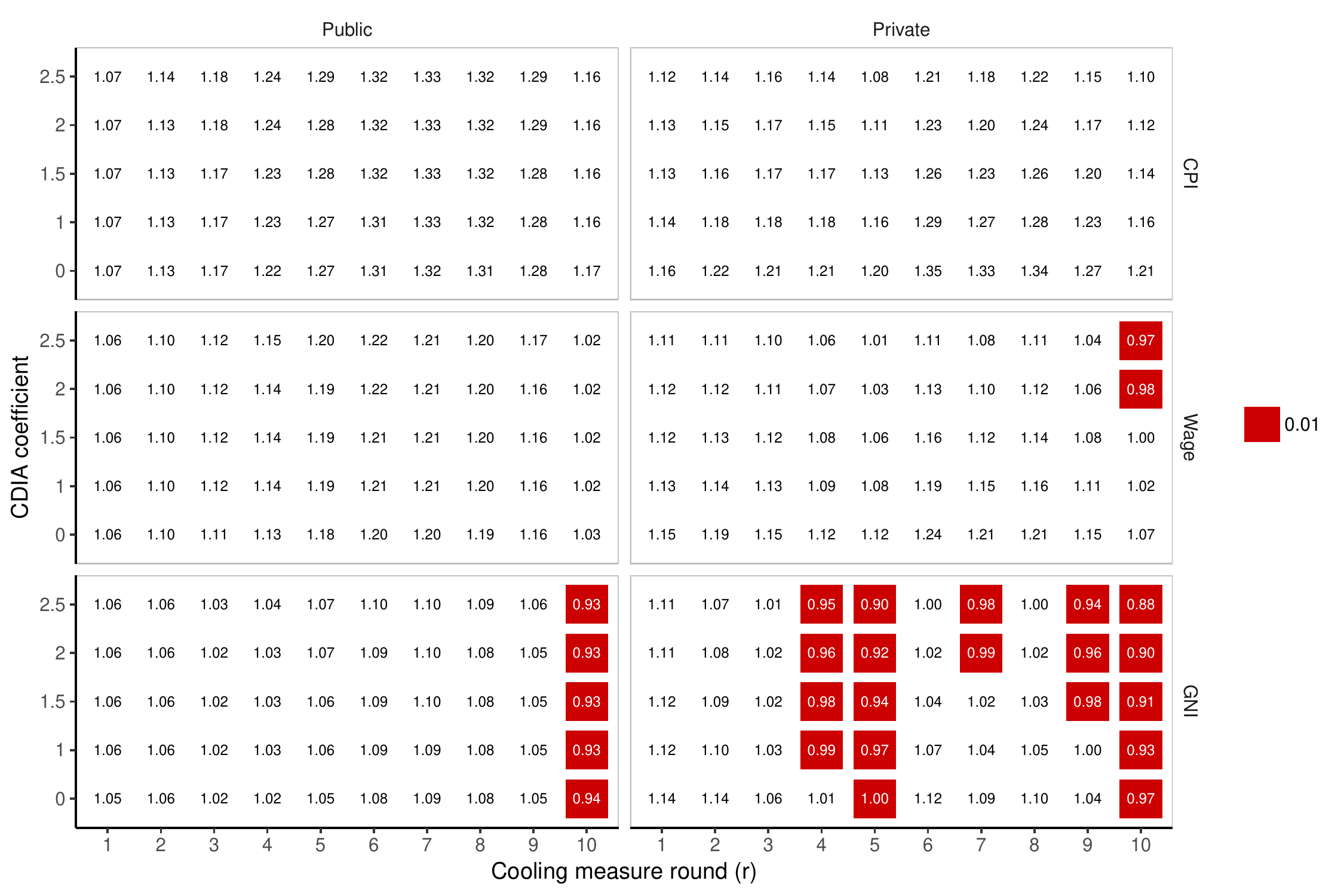}
\caption{{\bf Ratios of equivalent housing wealth for different cooling rounds.} The ratios relate the estimated equivalent wealth for round $r$ to the equivalent wealth estimated for the base period 0. CDIA is the constant degree  of inequality aversion $\nu$, see \autoref{eq:equivalent_wealth}. The p-values are for the null hypothesis that the respective ratio is larger than one. The p-values are based on bootstrapped values that are rounded up, whenever applicable, to the closest of the usual significance levels $(1\%,5\%,10\%)$. Bootstrap is conducted under the null hypothesis and uses $B=1000$ replications.}
\label{fig:equivalent_wealth}
\end{figure}

\newpage

\begin{figure}[htp!]
\centering
\includegraphics[width = \textwidth]{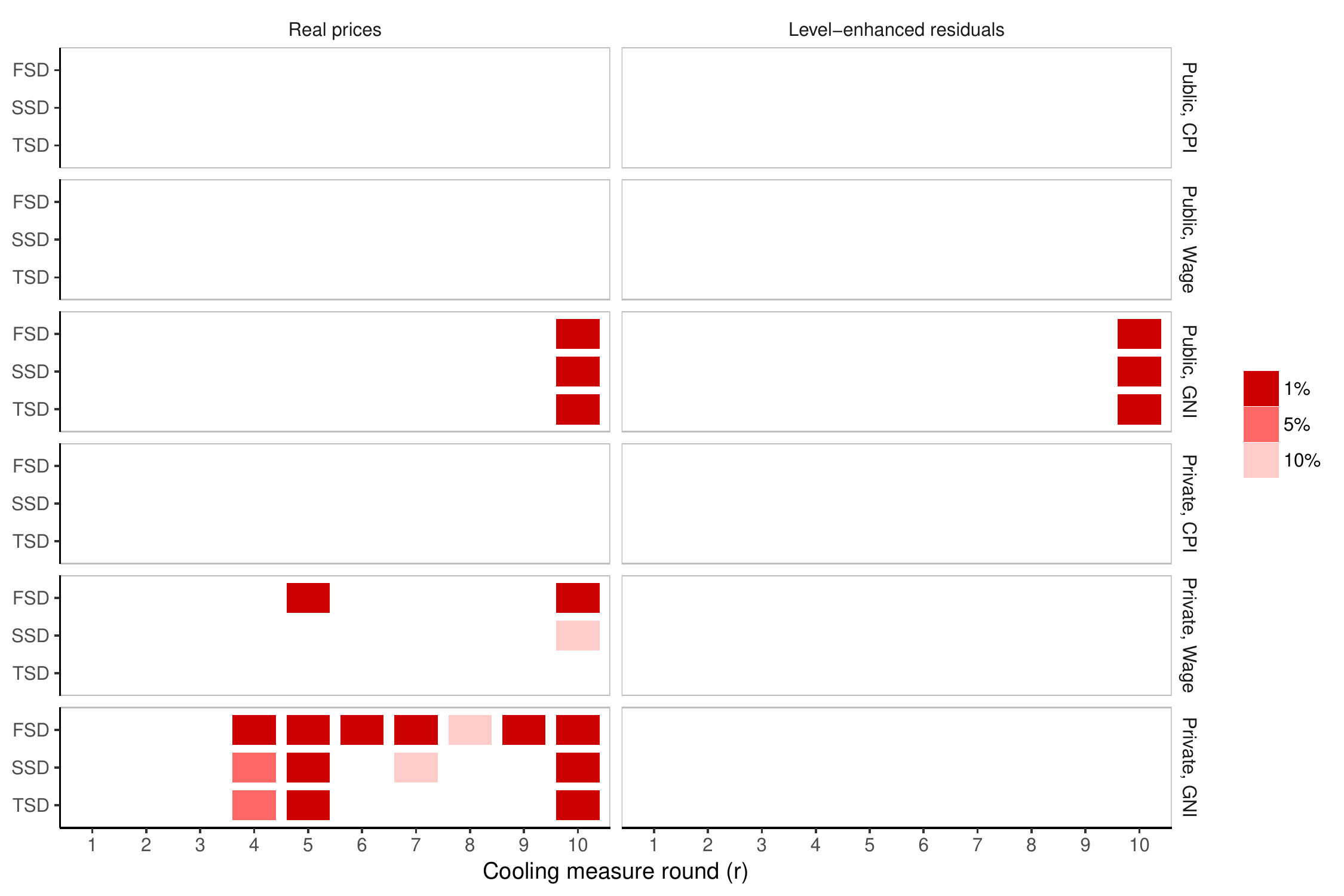}
\caption{{\bf Tests of stochastic dominance of price distributions for each cooling round relative to the base period.} Shows the p-values for the null hypothesis that the housing wealth distribution $F_r$ dominates $F_0$, for $r\geqslant1$. The p-values are based on bootstrapped values that are rounded up, whenever applicable, to the closest of the usual significance levels $(1\%,5\%,10\%)$. Bootstrap is conducted under the null hypothesis and uses $B=1000$ replications.}
\label{fig:SD_hypothesis1}
\end{figure}

\newpage

\begin{figure}[htp!]
\centering
\includegraphics[width = \textwidth]{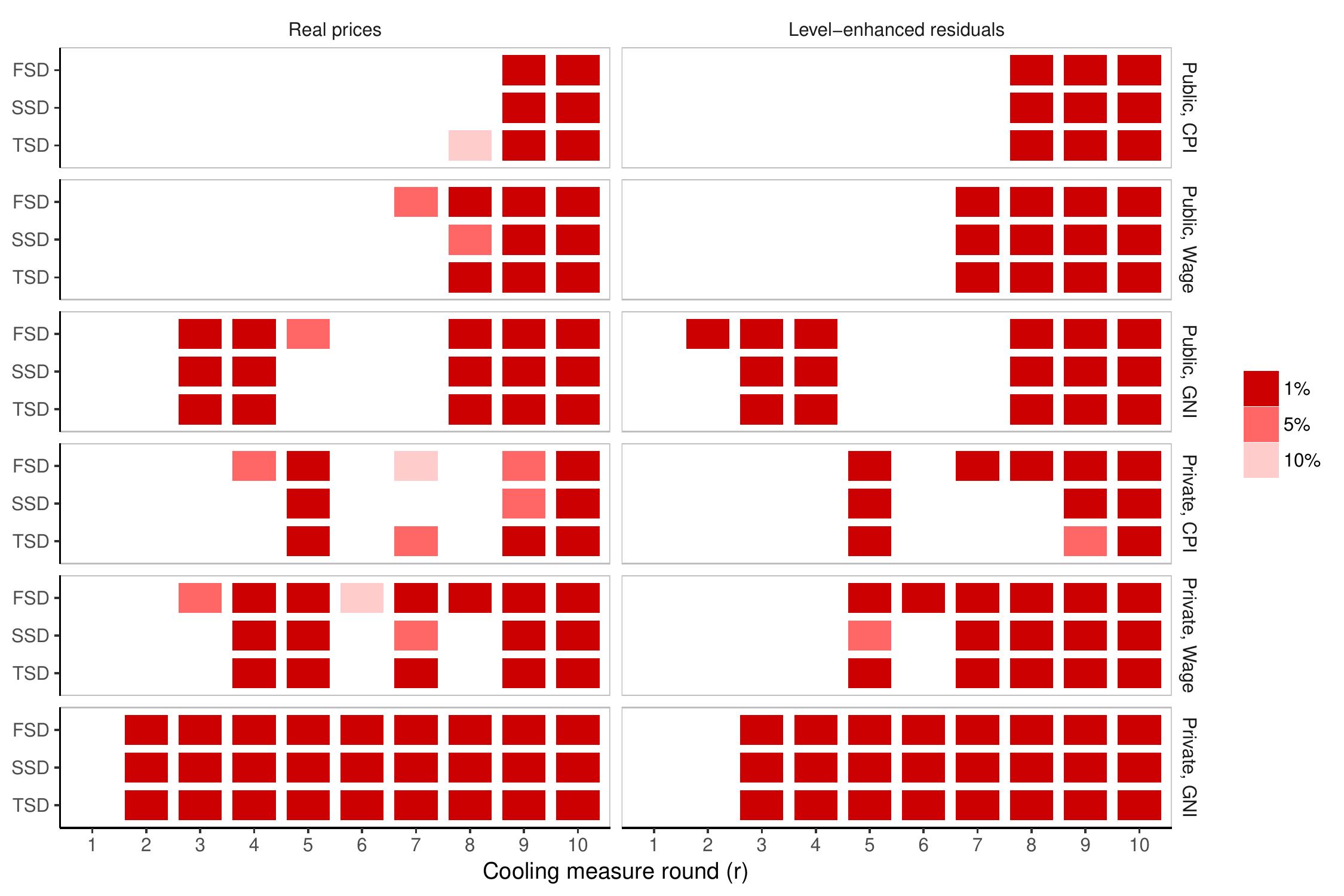}
\caption{{\bf Tests of stochastic dominance of price distributions for each cooling round relative to all preceding rounds.} Shows the p-values for the null hypothesis that the housing wealth distribution $F_r$ dominates stochastically all $F_k$ from previous rounds, $r>k\geqslant0$. The p-values are based on bootstrapped values that are rounded up, whenever applicable, to the closest of the usual significance levels $(1\%,5\%,10\%)$.}
\label{fig:SD_hypothesis2}
\end{figure}

\newpage

\begin{figure}
\centering
  \centering
\includegraphics[width = \textwidth]{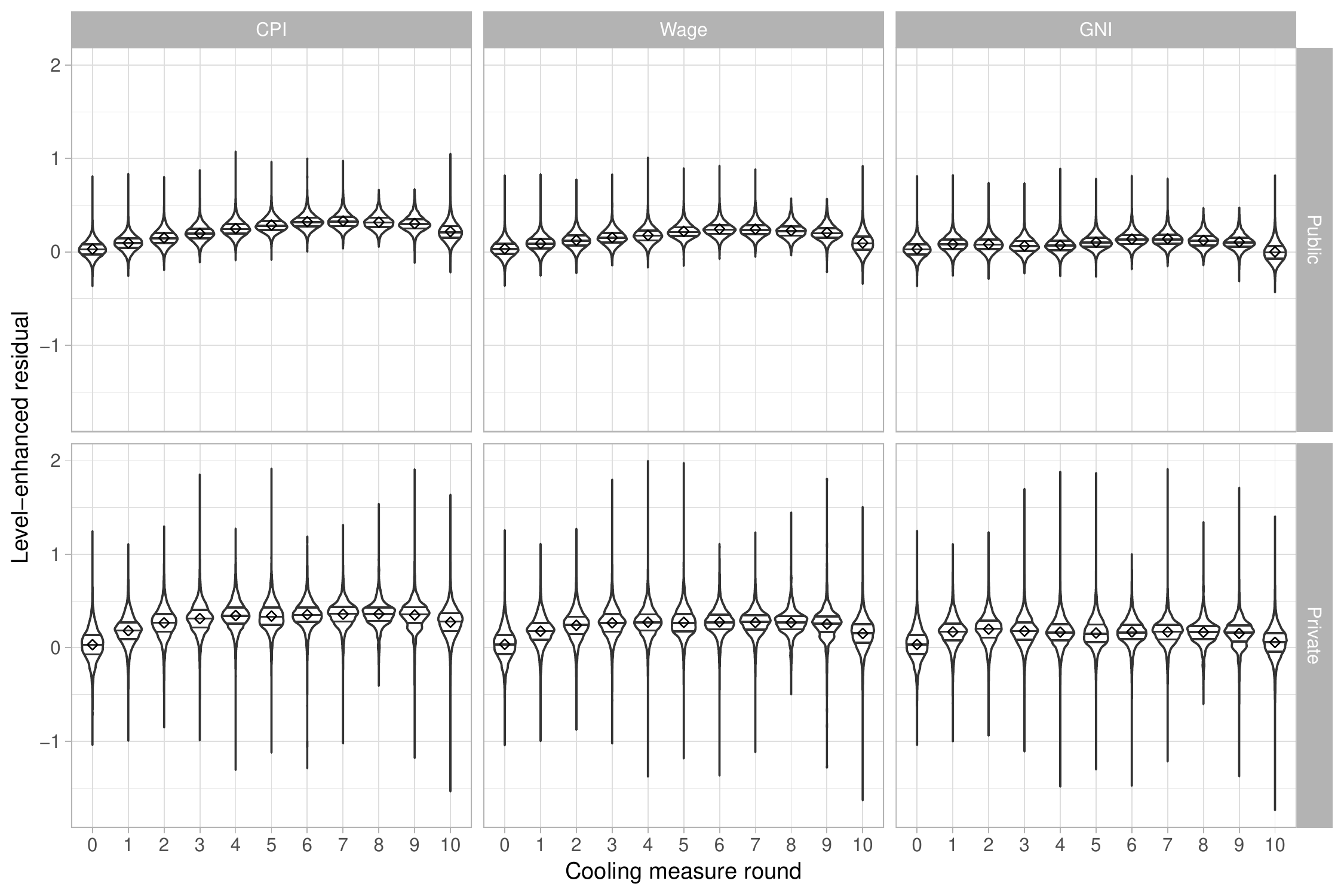}
 \caption{{\bf Distribution of level-enhanced residuals by cooling measure round.} Shows violin plots for the distributions of level-enhanced residuals in the public and the private  sector for the different cooling measure rounds. Horizontal lines in the violin plots are interquartile range and median, symmetric vertical lines are estimated kernel densities. The diamond indicates the arithmetic average. Round 0 corresponds to the base period.}
\label{fig:violin_level_enhanced_residuals}
\end{figure}

\end{document}

%% file: cooling_introduction.tex
\section{Introduction}

The financial crisis in 2007/08 has revealed that a regulatory regime which focuses solely on the solvency of individual financial institutions cannot ensure financial stability. Triggered by a correction of excessive house price growth, institutions overexposed to the housing sector experienced severe funding problems. Reliance on short term funding and fire sales of mortgage-related assets intensified these problems.
The interconnected financial obligations between institutions implied eventually that the insolvency of a few resulted in a seize-up of the whole financial system \citep{brunnermeier09,hansonetal11}.
Based on this experience, financial regulators in many countries have introduced  macroprudential measures to reduce systemic risk \citep{boe11,galatimoessner13}. 

Measures to control house price growth directly---so-called \emph{cooling measures}---have become popular in advanced Asian countries, such as Hong Kong and Singapore, but also in Europe \citep{darbarwu15}. When growth is excessive, tighter loan-to-value (LTV) and  debt-service-to-income (DTI) limits should curtail rising house prices.Similarly, sales and capital gain taxes will increase transaction costs, which makes speculative purchases less attractive. Empirical studies have found that cooling measures are indeed effective \citep{ceruttietal17,kuttnershim16,zhangzoli16}. 

No paper has examined yet the consequences of cooling measures on the distribution of housing wealth. For instance, a reduction of the average house price can imply that values of all dwellings fall by a small amount or that values  of small dwellings fall by a substantial amount. As smaller dwellings are owned dominantly by poorer households, the latter puts more burden on them.  If policymakers care about how housing wealth is distributed, they will not be indifferent to such consequences of cooling measures.

Our paper is the first that examines effects of cooling measures on the distribution of housing wealth. We use three related approaches to assess the distributional effects of cooling measures on social welfare and apply them to Singapore.
The first approach uses an explicit parametric welfare function with a constant degree of inequality aversion. While the choice of the inequality aversion parameter provides some flexibility for the welfare assessment, the first approach is fairly restrictive. The second approach places only mild restrictions on the form of welfare function and we apply tests of stochastic dominance up to order three the housing wealth distribution. Both the first and the second approach use transaction data and we apply re-weighting to make then representative. The third approach examines the expected utility of the owner of the average home who plans to sell. We apply a monotone price transformation that is driven by a systematic trend and unsystematic transaction noise. Cooling measures  will affect both the price trend and the unsystematic component. 

Singapore is well suited for an examination of the distributional effects of cooling measures. Starting in September 2009, cooling measures were implemented with increasing intensity over ten rounds until the full framework was in place in December 2013.\footnote{There have been a few minor adjustments to the regulatory framework since then.} Our data cover 2009Q1-2016Q2 and allow to assess the total effect on social welfare once all cooling measures are in place, but allow also to examine the changes in welfare once new measure are introduced. An additional source of comparative evidence comes from the segmentation of the housing market in Singapore. While the public sector is under state control and caters to citizens and permanent residents, the private sector offers housing to foreigners and well-off citizens. The private sector has been the  main target of cooling measures and the authorities were keen to avoid spillovers of cooling measures from the private into the public sector.

In this paper, we provide three novel pieces of evidence on the dynamics of housing wealth distributions amid cooling measures in Singapore. Firstly, the housing wealth in both the public had generally remained above the post-crisis period until the tenth round of cooling measures. Whereas, owners in the private sectors experienced deteriorating housing wealth from the fourth round. Secondly, the housing wealth did not change monotonically across the cooling measure rounds. Particularly, housing wealth in the private sector was adversely affected slightly earlier as compared to that in the public sector, due to that the initial rounds targeted at the private sector only. Thirdly, upon examining the housing wealth of a notional household owning a dwelling with standard characteristics, we find that this household enjoys better housing wealth than the post-crisis period even in the private sector. 

The rest of the paper is organised as follows. \autoref{sec:context} discusses the system of housing provision in Singapore and details the different cooling measures that have been implemented. \autoref{sec:methodology} presents the statistical methodology.  \autoref{sec:data} discusses the data and \autoref{sec:results} presents the results.  \autoref{sec:conclusion} concludes. The Appendix contains further details on the methodology.

%% file: cooling_context.tex
\section{Housing in Singapore}\label{sec:context}

\subsection{Homeownership and wealth}\label{subsec:homeownership_wealth}

The promotion of home ownership is an important objective in Singapore. Policy makers anticipate that housing wealth makes owners responsible and hard-working citizens. To promote ownership, the majority of new flats are constructed by the public sector Housing and Development Board (HDB). These flats are then assigned to citizens on long leaseholds at prices below construction cost. The type of flat that can be bought is determined mostly by household's characteristics, such as marital status and family size. After a minimum occupation period, citizens can sell their HDB flats at market prices. Both citizens and permanent residents can engage in the resale market for HDB flats.

Policy makers regard private housing as necessary complement to public provision. The share of private sector housing to all housing has increased steadily over the 2009-2016 period from twenty-two to twenty-six percent  \citep{statyearbook16,statyearbook17}. In the private sector, commercial development companies provide mostly top-end properties, such as single-family houses and exclusive apartments. Foreigners are not entitled to HDB flats and buy in the private sector as do well-off citizens who can afford top-end dwellings. Citizens finance HDB flat purchases with savings accumulated in the obligatory pension scheme CPF and with loans from the HDB or from private sector financial institutions, such as banks and development companies. Buyers in the private sector can use only funding from financial institutions.

\begin{center}
 [\autoref{tab:descriptives_context} about here.]
\end{center}

\autoref{tab:descriptives_context} (Panel A)  shows that policymakers have been very successful in generating ``a home-owning society'' \citep[p.21]{phang07}. Over the 2009-2016 period, ninety percent of all households in Singapore own their home. If we discard the small segment of one- and two-room flats, occupied mainly by households in need, then ninety-five percent of households in public sector dwellings are homeowners.  The home ownership rate in the private sector is smaller, but with eighty-three percent  still high  compared to many other developed countries. The high level of home
ownership is reflected both in households' gross wealth and liabilities. Residential real estate accounts for nearly half of all assets, see
\autoref{tab:descriptives_context} (Panel B).  Mortgages dominate by far liabilities, see \autoref{tab:descriptives_context} (Panel C).  Mortgages are underwritten dominantly by financial institutions,  with a volume increasing from 67\% in 2009 to  83\% in 2016. Compared to other developed countries, the exposure of household wealth to real estate  is high in Singapore  \citep{guisoetal02}.

\subsection{Cooling measures}

In the wake of the financial crisis 2007/08, the Monetary Authority Singapore (MAS) recognised ``that too much reliance had been placed [...] on financial institutions themselves to manage risk and on markets to be self-correcting'' \citep[p.9]{MAS_TER}. To limit  ``build-up of risks'' caused by  excessive credit growth and the ``concentrated exposures to particular sectors or asset classes'', the MAS decided to use macroprudential measures whenever necessary in the future  \citep[p.26]{MAS_TER}.

The economy in Singapore recovered quickly after the crisis and a risk build-up in the housing market was diagnosed as early as 2009. The MAS, supported by other government agencies, started in the same year to implement cooling measures. These should ``(i) promote a stable and sustainable property market where prices move in line with economic fundamentals, (ii) encourage greater financial prudence among property purchasers, and (iii) maintain sound lending standards.'' \citep[Box I]{masfsr_11}. Ten rounds of cooling measures are implemented by the end of 2013; the implementation dates are given in \autoref{tab:n_obs}. Since then, only slight amendments to the regulatory framework have been made.

Rounds one to six target mostly the private sector. \emph{Round 1} releases land to increase private development and completed dwellings must be sold sooner to avoid fines. This should increase supply.  To curb demand, loan facilities for the acquisition of uncompleted dwellings are removed.  \emph{Round 2} releases more land for private development. To curb speculative demand, a seller stamp duty (SSD) must be paid if a dwelling is owned  less than a year. The loan to value ratio (LTV) for bank mortgage loans is capped at 80\%. To curb demand further, \emph{Round 3} extends the SSD to sales of dwellings owned for less than three years. The LTV is capped at 70\%. Borrowers applying for additional loans have to pay a higher cash deposit. \emph{Round 4} extends the SSD to sales of dwellings owned less than four years. SSD tax rates are raised substantially.  The SSD does not affect homeowners in the public sector directly, as they can own only one dwelling. \emph{Round 5}  introduces an additional buyer stamp duty (ABSD) that has to be paid in all transactions where the buyer is either a foreigner, a permanent resident owning already a dwelling, or a citizen who owns already more than one dwellings. Citizens and permanent residents in the public sector are not affected directly, as they can own only one dwelling.
\emph{Round 6} tightens covenants of private mortgage loan  by capping maturities to 35 years and reducing LTVs for loans with shorter maturities. This can impact on the HDB resale market, because mortgage loans from financial institutions are used for the purchase of such dwellings.

Rounds seven to nine target both housing sectors. \emph{Round 7} extends and tightens the ABSD.  Foreigners pay now 15\%, permanent residents pay 5\% for their first dwelling (10\% for consecutive), and citizens pay 7\% for their second dwelling (10\% for consecutive). The tightening has no direct impact on citizens in the HBD market, but permanent residents are affected if they want to purchase in the HDB market. Permanent residents are  also no longer allowed to sublet HDB flats. LTVs for additional housing loans are  capped and require  a higher minimum cash down payment.  The mortgage service ratio (MSR) of loans for new and resale HDB flats is lowered, although to a lesser extend if the HDB underwrites the loan. Given gross income, a lower MSR means that the potential  loan size is reduced. The use of CPF savings for the purchase of HDB resale flats becomes more difficult.
\emph{Round 8}  sets new rules for loan underwriting.  An applicant's total debt service ratio (TDSR) has to consider all outstanding liabilities and cannot be larger than 60\%. \emph{Round 9} targets the HDB resale market and reduces the maximal maturity of HDB loans from 30 to 25 years and reduces  the MSR by five percentage points to 30\%. The maximal maturity of bank loans for the purchase of HDB flats is reduced from 35 to 30 years. These measures reduce credit availability for purchases in the HDB resale sector. Permanent residents have to wait longer before they can enter the HDB resale market. \emph{Round 10} to \emph{Round 12} bring only immaterial adjustments to the cooling measures already in place.

Up to the end of 2016, the cooling measures in Sinagore have successfully prevented a systemic crisis. However, this does not mean that undesirable effects are absent. Targeting house price and loan growth will have an effect on the housing wealth distribution. As we discussed at the beginning of this section, housing wealth is central to Singapore's aspiration to a fair participation of its residents in the country's economic success. To be able to assess cooling measures fully, we must also look at the  welfare of those who own dwellings.
We present next the methodology we use for this task.

%% file: cooling_methodology.tex
\section{Empirical methodology}\label{sec:methodology}

\subsection{Welfare assessment}\label{subsec:welfare_assessment}

The cumulative distribution function $F_i(p)$ of gross housing wealth $p$ in $i$---a combination of housing market segment and cooling round---has support $[0,\overline{p}_i]$, with $F_i(0)=0$ and $F_i(p)=1$ for $p\geqslant\overline{p}_i$. The corresponding density function is $f_i(p)$. Housing market segment refers to either the public and or the private sector. The Singaporean state cares about the housing wealth of resident households, we model this with the utilitarian social welfare function
\begin{equation}\label{eq:welfare_function} W_i=\int_{0}^{\overline{p}_i}u(p)f_i(p)dp
\end{equation}
The welfare function aggregates household utility $u(p)$ over all homeowners. The welfare function treats equals equally and is compatible with the Pareto criterion, because welfare increases strictly monotonically if one household receives more utility while the utilities of all others remain unchanged.

The function in \autoref{eq:welfare_function} does not provide enough structure as such to rank wealth distributions. A particular choice of $u(p)$ would make such rankings possible.
For instance, we would be able to determine the equivalent housing wealth $e_i$, defined implicitly through
\begin{equation}\label{eq:equivalent_wealth_definition}
 \int_{0}^{\bar{p}}u(e_i)f_i(p)dp=\int_{0}^{\bar{p}}u(p)f_i(p)dp
\end{equation}
and explicitly through $e_i=u^{-1}(W_i)$. If all households had the level $e_i$ of housing wealth, social welfare would be same as under the current---possibly unequal---wealth distribution $F_i$.\footnote{ Conceptually, $e_i$ is identical to the equally distributed equivalent level of income introduced by \citet{atkinson70}.} Given  $F_i$ and $F_j$, $i\neq j$, the distributions can be ranked with the equivalent wealth ratio $\psi_{ji}\equiv e_j/e_i$. If $\psi_{ji}>1$, welfare under $F_j$ is higher than under $F_i$.

A particular choice of $u(p)$ brings normative welfare trade-offs that might not be shared by everyone. We will therefore conduct our analysis also by testing for stochastic dominance (SD) and  will place up to three economically meaningful and widely agreeable shape restrictions on the utility functions. Let $\mathcal{U}_s$ denote sets that collect all utility functions that share particular shape restrictions, $s=1,2,3$. As we will see below, the shape  restrictions tighten with $s$, so that  $\mathcal{U}_3\subset\mathcal{U}_2\subset\mathcal{U}_1$. The following general theorem holds \citep{bawa75}: Given two wealth distributions $F_j$ and $F_i$, $i\neq j$, if welfare under $F_j$ is higher than under $F_i$ for all $u\in\mathcal{U}_s$, then $F_j$ must dominate $F_i$ stochastically at order $s$ (necessary condition). If $F_j$ dominates $F_i$ stochastically at order $s$, then welfare is higher under $F_j$ than under $F_i$ for all $u\in\mathcal{U}_s$ (sufficient condition). However, SD is only a partial ordering and it might be possible that some distributions cannot be ranked. In such a case, a welfare ranking requires a welfare function. 
 
We discuss now the sufficient conditions and the role of $\mathcal{U}_s$. Each of the two distributions $F_l(p)$, $l=i,j$, have support $[0,\max(\overline{p}_i,\overline{p}_j)]$ with $F_l(0)=0$ and $F_l(p)=1$ for $\overline{p}_l<p\leqslant\overline{p}$.
We define $D^{(0)}_{ji}(p)\equiv f_j(p)-f_i(p)$ and for $s\in\{1,2,3\}$
\begin{equation}
 D^{(s)}_{ji}(p)\equiv\int_0^pD^{(s-1)}_{ji}(z)dz \label{eqn:D}
\end{equation}
Note that $dD^{(s)}_{ji}(p)/dp=D^{(s-1)}_{ji}(p)$.

The difference in social welfare generated by the two distributions is
\begin{equation}\label{eq:FSD_condition_1}
 W_j-W_i = \int_0^{\overline{p}}u(p)D^{(0)}_{ji}(p)dp
\end{equation}
Integration by parts of the right-hand side of \autoref{eq:FSD_condition_1} leads to
\begin{equation}\label{eq:eq:FSD_condition_2}
  W_j-W_i=
  -\int_0^{\overline{p}}u'(p)D^{(1)}_{ji}(p)dp
\end{equation}
where we use that $D^{(1)}_{ji}(\overline{p})=D^{(1)}_{ji}(0)=0$. 
If the  utility function comes from the set $\mathcal{U}_1=\{u|u'>0\}$, then a sufficient condition for $W_j-W_i>0$ is
\begin{equation}\label{eq:fsd_condition}
 D^{(1)}_{ji}(p)=F_j(p)-F_i(p)\leqslant0
\end{equation}
with the inequality strict for some $p\in[0,\overline{p}]$. The condition in \autoref{eq:fsd_condition} corresponds to \emph{first-order stochastic dominance} (FSD) of $F_j(p)$ over $F_i(p)$. Note that FSD implies $\E[p|F_j]\geqslant\E[p|F_i]$.\footnote{Integration of \autoref{eq:fsd_condition} over the full support and using integration by parts leads to this result.} The shape restriction $u'(p)>0$ implies that utility increases strictly in housing wealth, which seems uncontroversial. The shape restriction ensures that the social welfare function in \autoref{eq:welfare_function} fulfils the Pareto criterion with respect to housing wealth. While a weak assumption on $u(p)$, FSD is a strong requirement for two distribution functions.

Integration by parts of the right-hand side of \autoref{eq:eq:FSD_condition_2} leads to
\begin{equation}\label{eq:SSD_condition}
 W_j-W_i =-u'(\overline{p})D^{(2)}_{ji}(\overline{p})+\int_0^{\overline{p}}u''(p)D^{(2)}_{ji}(p)dp
\end{equation}
where we use that $D^{(2)}_{ji}(0)=0$. If the utility function comes from the set $\mathcal{U}_2=\{u|u'>0,u''<0\}$, then a sufficient condition for $W_j-W_i>0$ is
\begin{equation}\label{eq:ssd_condition}
D^{(2)}_{ji}(p)=\int_0^{p}\left\{F_j(z)-F_i(z)\right\}dz\leqslant0
\end{equation}
with strict inequality for some $p\in[0,\overline{p}]$. The condition in \autoref{eq:ssd_condition} corresponds to \emph{second-order stochastic dominance} (SSD) of $F_j(p)$ over $F_i(p)$. Note that SSD implies $\E[p|F_j]\geqslant\E[p|F_i]$.\footnote{This follows immediately from \autoref{eq:ssd_condition} at $p=\overline{p}$ and integration by parts.} The second shape restriction $u''(p)<0$ implies \emph{transfer sensitivity} \citep{atkinson70}. A transfer of wealth from an owner with a high value dwelling to an owner of a low value dwelling increases social welfare. If this implication is accepted, then SSD can be used to rank distributions. FSD implies SSD, but not the other way around. SSD is a weaker criterion than FSD.

Integration by parts of the second term on the right-hand side of \autoref{eq:SSD_condition} leads to
\begin{equation}\label{eq:TSD_condition}
 W_j-W_i =-u'(\overline{p})D^{(2)}_{ji}(\overline{p})+u''(\overline{p})D^{(3)}_{ji}(\overline{p})-\int_0^{\overline{p}}u'''(p)D^{(3)}_{ji}(p)dp
\end{equation}
where we use that $D^{(3)}_{ji}(0)=0$. If the utility function comes from the set $\mathcal{U}_3=\{u|u'>0,u''<0,u'''>0\}$, then 
\begin{subequations}\label{eq:tsd_conditions}
\begin{equation}\label{eq:tsd:integral_condition}
\begin{aligned}
D^{(3)}_{ji}(p)&=\int_0^pD^{(2)}_{ji}(p)dp\\
&=\int_0^p\int_0^y\left\{F_j(z)-F_i(z)\right\}dzdp\leqslant0
\end{aligned}
\end{equation}
with strict inequality in \autoref{eq:tsd:integral_condition} for some values in $p\in[0,\overline{p}]$ and
\begin{equation}\label{eq:tsd:mean_condition}
D^{(2)}_{ji}(\overline{p})\leqslant 0
\end{equation}
\end{subequations}
are sufficient conditions for $W_j-W_i>0$. The condition in \autoref{eq:tsd:mean_condition} is equivalent to $\E[p|F_j]\geqslant\E[p|F_i]$, as can be derived with integration by parts. The conditions in \autoref{eq:tsd_conditions} correspond to \emph{third-order stochastic dominance} (TSD) of $F_j(p)$ over $F_i(p)$. The third share restriction $u'''(x)>0$ implies \emph{diminishing transfer sensitivity}, which means that progressive transfers at the lower end of the wealth distribution receive more weight than such transfers at the upper end \citep{atkinson08,shorrocksforster87}. If this implication is accepted, then TSD can be used to rank distributions. SSD implies TSD, which is a weaker criterion than both FSD and SSD.

\subsection{Empirical implementation}

\subsubsection{Preliminary steps}

To assess empirically the welfare impacts of the different rounds of cooling measures in the public and the private sector of Singapore's housing market, we need estimates of the dwelling value distributions $F_i(p)$ separately for each of the two different housing segments for each of the ten cooling rounds and for a base period. If transaction prices reflected these distributions, we could use them directly in our analysis. However, it is possible that some dwellings types are transacted more or less frequently than it corresponds to the composition of the owner-occupied housing stock. 

To deal with this possibility, we apply post-stratification in the first two empirical approaches. In particular, we re-weight the transactions in our estimators based on extra-sample information on the actual type composition of the owner-occupied stock. This results in the distribution of dwelling values.
The re-weighting steps are analogous for the two housing sectors---public and private---and we concentrate on one sector to avoid excessive indexing. For each dwelling type $t\in\{1,\hdots,T\}$ and cooling round $r\in\{0,\hdots,10\}$, we compute the weights
\begin{equation}\label{eq:weights}
 w^{t}_r = \frac{S_r^t}{S_r}\left(\frac{N^t_r}{N_r}\right)^{-1}
\end{equation}
The ratio in the numerator on the right-hand side relates the stock of owner-occupied dwellings of type $t$, $S^t_t$, to the total stock $S_r$ of owner-occupied dwellings in round $r$. The ratio in the denominator does the same for number of observed transactions. If relatively more dwellings of type $t$ are transacted than are in the stock, then the weight is below one. We note further that for all transactions $k$ observed for round $r$ and each type $t$
\begin{equation}\label{eq:reweighting_effect}
 w^t_r\sum_{k=1}^{N_r}\mathbf{1}(k,r;t)=\frac{S^t_r}{S_r}N_r
\end{equation}
where the indicator function $\mathbf{1}(k,r;t)$ becomes one if dwelling $k$ transacted in round $r$ is of type $t$, it becomes zero otherwise. \autoref{eq:reweighting_effect} shows that re-weighting ensures that the effective sample size for each dwelling type conforms with the size expected from the owner-occupied housing stock. Note that summing \autoref{eq:reweighting_effect} over all dwelling types gives $N_r$.

Cooling measures affect prices through new supply, which changes the composition of the housing stock, and through new financial regulations,  which changes the composition of households who can  participate in the market. This alters housing supply and demand and has consequences for the general price trend. In addition to this systematic trend effect, cooling measures will also affect the unsystematic noise that accompanies each transaction. 
We examine this using level-enhanced residuals. The residuals come from the familiar log-linear hedonic regression model. To be specidic, we fit the partial-linear model
\begin{equation}\label{eqn:hedonic}
y_{k,r} = \vek{z}_{k,r}\bsy{\delta}_r + \vek{s}_{k,r}\bsy{\beta}_r + g_r(\vek{l}_{k,r}) + \epsilon_{k,r}
\end{equation}
separately for each round $r$ and each of the two housing sectors. $y_{k,r}$ is the log real price of dwelling $k$, one of the $N_r$ transactions in round $r$. We denote with $Q$ and $Q_r$ the number of quarters in our sample period and the number of quarters that fall within round $r$, respectively. The $1\times(Q_r-1)$ vector $\vek{z}_{k,r}$ has a one as first (second) entry if $k$ is transacted in the second (third) quarter of round $r$ and so forth. The remaining elements are all zero.  The coefficients in $\bsy{\delta}_r$ consider therefore the quarterly average price level relative to the first quarter within round $r$.
The $1\times C$ vector $\vek{s}_{k,r}$ collects dwelling's structural characteristics, such as dwelling type and floor area. The $C\times 1$ vector $\bsy{\theta}_r$ contains the implicit prices for these characteristics. The smooth $g_r(\cdot)$ is a function dwelling's two geo-location coordinates given in the vector $\vek{l}_{k,r}$. It captures local amenity effects in a flexible manner. We model $g_r(\cdot)$ with splines and fit \autoref{eqn:hedonic} using a penalised splines estimator \citep{ruppertetal03,wandetal05}.

Once the regression residuals are estimated, we compute the level-enhanced residuals 
\begin{equation}\label{eq:level_enhanced_residual}
 \hat{p}_{k,r}\equiv\tilde{\vek{z}}_{k,r}\vek{i}+\hat{\epsilon}_{k,r}
\end{equation}
where $\vek{i}$ is a $Q\times 1$ vector that contains the log of a  quarterly house price index. We use official price indices converted into real terms with the same deflators as the transaction prices and normalised to be one in the first quarter of our sample period.  The $1\times Q$ vector $\tilde{\vek{z}}_{k,r}$ has a one in the column corresponding to the quarter in which $k$ is transacted and zeros elsewhere. The first term on the right-hand side of \autoref{eq:level_enhanced_residual} thus picks the log price level for the respective quarter and the second term adds the residual to it. The level-enhanced residuals consider therefore systematic trend and unsystematic transaction noise. The notional owner of the standard home (which is underlying the official price index) faces this distribution if she wants to sell in $r$.

\subsubsection{Equivalent wealth ratio tests}\label{subsec:welfare_equivalent_housing_wealth}

Computation of $W_i$  and $e_i$ requires a functional form for $u(p)$. We choose
\begin{equation}\label{eq:functional_form_u}
 u(p)=\begin{cases} \dfrac{p^{1-\nu}}{1-\nu} & \mbox{for $\nu\geqslant0$ and $\nu\neq 1$}\\
 \ln p & \mbox{for $\nu=1$}
 \end{cases}
\end{equation}
which has been introduced for inequality measurement by \citet{atkinson70}. The utility function in \autoref{eq:functional_form_u} is element of $\mathcal{U}_1$ for $\nu=0$ and of $\mathcal{U}_3$ for $\nu>0$. Using \autoref{eq:functional_form_u} in \autoref{eq:equivalent_wealth_definition} gives
\begin{equation}\label{eq:equivalent_wealth}
 e_{i} = \begin{cases} (1-\nu)W_i^{1/(1-\nu)} & \mbox{for $\nu\geqslant0$ and $\nu\neq 1$}\\
 \exp\{W_i\} & \mbox{for $\nu=1$}
 \end{cases}
\end{equation}
The particular values for $W_i$ and $e_i$  depend  on the chosen degree of inequality aversion $\nu$. For instance, if $\nu=0$, $e_i$ corresponds to the average wealth and no attention is given to  distributional aspects. The larger $\nu$, the more relevant become distributional aspects.

We focus our analysis on the equivalent wealth $\psi_{r0}$ that compares round $r\geqslant1$  with the base round. If $\psi_{r0}<1$, then social welfare in round $r$ is lower than in the base period. This could be either because the average wealth level is lower, or because wealth inequality has increased or both. Such an outcome should concern policy makers. $\psi_{r0}\geqslant1$ will give no reason for concern. We will test the  the hypotheses $\vek{H}_0\!:\; \psi_{r0}\geqslant1$ versus $\vek{H}_1\!:\; \psi_{r0}<1$ with the test statistic 
\begin{equation}\label{eq:test_statistic_equivalent_wealth}
\theta_{r0}\equiv \frac{\hat{\psi}_{r0}-1}{\hat{\sigma}_{\hat{\psi}_{r0}}}
\end{equation}
 \autoref{app:testing_equivalent_wealth} gives details on estimation and inference. If we can reject the null for a given significance level, then this indicates that  welfare as measured by the particular welfare function in round $r$ has deteriorated relative to the base period.

\subsubsection{Dominance tests}\label{subsec:statistical_implementation_sd}

We test for stochastic dominance using extensions of the Kolmogorov-Smirnov test   \citep{klecanetal91,barrettdonald03,lintonetal05}. The test statistic is
\begin{equation}\label{eqn:d*}
d^{(s)}_{j} \equiv \max_{i \neq j} \sup_{p \in \left[0, \bar{p}\right]} \left[D_{ji}^{(s)} \left( p \right)\right]
\end{equation}
where $D_{ji}^{(s)}\left(p \right)$ is defined in \autoref{eqn:D}.
The max operator becomes relevant whenever we compare the distribution from round $j$ with several distributions from preceding rounds simultaneously. 
The term with the sup operator in \autoref{eqn:d*}  will be at most zero if $j$ dominates the distribution from round $i$ stochastically at order $s$. It will be negative if the dominance is strict for all $p\in[0,\overline{p}]$. A positive $d^{(s)}_{j}$, on the other hand, implies that there is no stochastic dominance of $j$ over $i$.  The hypotheses of the tests for stochastic dominance are thus $
\vek{H}_0\!\!:\; d^{(s)}_{j} \leqslant 0$ versus $\vek{H}_1\!\!: \; d^{(s)}_{j} > 0$.
If dominance of distribution $j$ is tested against distributions from multiple other rounds, $d^{(s)}_{j}$ becomes the maximum of the pairwise comparisons, as this provides the least favourable evidence for the null hypothesis. If rejected, round $j$ does not dominate stochastically all preceding rounds. For instance, if $j$ is the period in which all cooling measures are in place, then rejection of the null implies that $j$ does not dominate some of the preceding distributions stochastically. 

Policy makers in Singapore aim at cooling the housing market without impairing the housing wealth distribution. If policy makers care about the Pareto criterion and (diminishing) transfer sensitivity, but are otherwise agnostic about the shape of $u(p)$, SD becomes a necessary condition for their aim of  $W_j>W_i$, for  $j>0$ and $i\in\{0,\hdots,j-1\}$.   
Rejection of SD will then cause concern, because it indicates that the measures implemented may have destroyed welfare somewhere along the way.

To implement the tests, we use the empirical counterpart of \autoref{eqn:d*}
\begin{equation}\label{eq:estimator_d_hat}
\hat{d}_j^{(s)} = \max_{i \neq j} \sup_{p \in \left[0, \bar{p}\right]} {\sqrt{N_{ji}}} \left[\hat{D}_{ji}^{(s)}(p)\right]
\end{equation}
with $N_{ji} \equiv (N_j N_i)(N_j + N_i)^{-1}$ and use the bootstrap for inference, see \autoref{app:testing_stochastic_dominance} for details.

In case of the level-enhanced residuals, the bootstrap procedure has to take into account that the original observations $\hat{p}_{k,r}$ are  estimated \citep{lintonetal05}. This calls for an additional step in which we use 
the set of residuals  $\{\hat{\epsilon}_{kr}\}_{k=1}^{N_r}$ from the regression \autoref{eqn:hedonic} applied to the original transaction data. For each bootstrap replication, we draw a random sample with replacement of size $N_r$ from the set of residuals and construct artificial log prices
\begin{equation}
y_{k,r}^b\equiv\vek{z}_{k,r}\hat{\bsy{\delta}}_r+\vek{x}_{k,r}\hat{\bsy{\beta}}_r + \hat{g}_r(\vek{l}_{k,r}) + \epsilon_{k,r}^b
\end{equation}
for $k=1,\hdots,N_r$. The artificial $y^b_{k,r}$ is effectively the predicted log price from the original regression plus a re-sampled residual as new noise term.  Running the regression from \autoref{eqn:hedonic} for the artificial observation sample $\{(y_{k,r}^b,\vek{z}_{k,r},\vek{x}_{k,r},\vek{l}_{k,r})\}_{k=1}^{N_r}$ gives a new set of residuals  and, using \autoref{eq:level_enhanced_residual}, a new set of level-enhanced residuals $\{\hat{p}^b_{k,r}\}_{k=1}^{N_r}$. We replicate this exercise $B$ times separately for the rounds we want to compare. The artificial samples of residuals take to role of the bootstrap samples for the estimation of the distribution of test statistic under the null as before. The only difference to the implementation above is that we do not re-weight, as the residual analysis is for the transacted dwellings.

%% file: cooling_data.tex
\section{Data}\label{sec:data}

Quarterly transaction price indices for the public and the private sector are provided by the HDB and the Urban Redevelopment Authority (URA), respectively. The indices are computed as weighted median prices---since 2015 quality-controlled prices---over different dwelling types. The indices inform the public and policy makers about the price behaviour in Singapore. Consequently, we assess cooling measures' effects on average price levels with these indices.

The transaction data comes from Singapore Real Estate Exchange (SRX), the country's leading consortium of real estate agents. The data set covers the period from 1 February 2009 to 30 June  2016. Each observation in the data set contains dwelling's transaction date and information on its characteristics, such as dwelling type, storey if part of a multi-storey building, floor area, and location coordinates. Whereas transacted HDB dwellings are always re-sale leaseholds, transacted private sector dwellings can be new and freeholds.
Relative to official total transaction figures from HDB and URA, our data cover 85.8\% (2009M2-2010M6) and 97.5\% (2011M1-2016M6) of public sector transactions and 95.4\% of private sector transactions (2009M2-2016M6).\footnote{The lower coverage for the public sector in the early period comes from  retrospective information collection by SRX and members who could not always accommodate requests for historical data.} \autoref{tab:n_obs} presents the number of transactions in our data set in total and split for the different rounds of cooling measures. The public sector stock is larger than the private, but less transactions take place.  This can be partly explained by the observations that new public dwellings  are not allocated through the market and that a minimum holding period of five years is enforced.
\begin{center}
[\autoref{tab:n_obs} about here.]
\end{center}
\autoref{tab:descriptive_type} gives summary statistics for the transaction data. Private sector dwellings are on average much more expensive than public dwellings. Single-family structures are only available in the private sector are the most expensive dwelling type. Dwellings in private sector multi-family structures are not necessarily larger than dwellings in public sector structures, but are often new, freeholds, and located more centrally as measured by the distance to the City Hall.\footnote{The Singaporian market distinguishes between apartments and condominiums. The only difference between the two categories is that apartments are in more densely developed settlements.} Single-family houses are available only in the private market. Compared to the other property types, single-family houses are predominatly owned as freehold. The coverage ratios in the bottom rows of \autoref{tab:descriptive_type} show that transactions in the public sector are in line with the public sector stock composition. In the private sector, multi-family dwellings are transacted more frequently than would be expected given the stock and single-family structures are transacted less frequently.\footnote{The ownership ratios for the cooling rounds, including the incomplete years 2009 and 2016, are calculated on a pro-rata basis from yearly series published by DOS.}
\begin{center}
[\autoref{tab:descriptive_type} about here.]
\end{center}

\autoref{fig:violin_nominal_prices} plots the distributions of nominal transaction prices for each round of cooling measures. In the public sector, the median prices increases in early rounds and decreases afterwards. The distribution in the lower bottom becomes bimodal too. The distributions in the private sector look fairly similar, in particular, there is not much variation of the nominal median price over the different rounds.
\begin{center}
[\autoref{fig:violin_nominal_prices} about here.]
\end{center}

We convert nominal into real prices with the following series obtained from CEIC: the monthly consumer price index excluding accommodation cost (CPI); the quarterly series of the quarter-end average monthly wage rate (WR); and the quarterly GNI per resident household (GNI).\footnote{While  nominal gross national income is available at a quarterly frequency, number of resident households is not. We interpolate the latter from the available  yearly figures.}  The CPI deflator expresses dwelling values in purchasing power for consumption goods, the WR deflator in  multiples of labour income, and the GNI deflator in multiples of Singapore's per household income. Given policymakers' aspiration that residents should participate in the growth of the economy through their housing wealth, the latter deflator seems of particular importance for our welfare assessment. Given the transaction date, each observation is deflated with the corresonding monthly or quarterly realisation of the repective deflator variable.

We obtain the weights in \autoref{eq:weights} by calculating the owner-occupied stock by dwelling type with the information on the dwelling stock and home-ownership rates given in \citet{statyearbook16} and \citet{statyearbook17}.  The  \citeauthor{statyearbook16} information distinguishes in the public four dwelling types  and two in the private sector, see the second row of \autoref{tab:descriptive_type}. The thus calculated owner-occupied stock by dwelling type are yearly figures and we interpolate linearly to obtain the appropriate figures for each of the cooling rounds.

%% file: cooling_results.tex
\section{Effects of cooling measures}\label{sec:results}

\subsection{Price trend and dwelling value distribution}

Cooling measures should prevent \emph{excessive} house price growth. While there is no clear definition of what  constitutes excessive growth, most would agree that it occurs when house prices grow faster than the rest of the economy. Over the period 2009Q1 to 2016Q2, 
Singapore's nominal GDP has been growing at 6.8\% p.a., with positive quarterly  year-on-year growth in all except the first two quarters.\footnote{Own calculations using the GDP at current market prices (M014461).} Over the same period, the official house price indices, shown in \autoref{fig:average_prices}, have grown  at similar magnitudes  of  4.1\% p.a. in the public and 4.6\% p.a. in the private sector. However,  different from economic growth, quarterly  year-on-year price growth has been positive only until the middle of the period.
\begin{center}
 [\autoref{fig:average_prices} about here.]
\end{center}
\autoref{fig:average_prices} shows that the differential trajectories of house prices and the economy affect the behaviour of the real prices that are relevant for the welfare assessment. Over the period, CPI-deflated prices grow  in the public (private) sector  at 2.6\% (3.1\%) p.a. This consists, however, of strong growth to the peak 4.5 years within the period at 7.6\%  (8.4\%) p.a.  and prices that fall over the rest of the period at  4.5\% (4.2\%) p.a.  The other two real price series exhibit a similar split behaviour. WR-deflated prices peak after 4.25 (2.75) years, having grown at  6.0\% (10.9\%) p.a., and fall  then at 6.2\% (4.3\%) p.a. GNI-deflated prices peak after 4.25 (1.5) years, having grown at 3.4\% (14.7\%) p.a., and fall then at  6.2\% (5.0\%). Over the period, WR-deflated and GNI-deflated prices grow at  0.5\% (1.0\%) p.a. and  -0.4\% (0.2\%) p.a., respectively.
Compared with the G7 countries, three have similar nominal and CPI-deflated house price growth rates and five have similar income-deflated house price growth rates for the 2009Q1-2016Q2 period.\footnote{Nominal and CPI-deflated: Canada (5.5\%, 4.1\%),  Germany (3.4\%, 2.2\%), UK (3.9\%, 2.1\%); disposable-income-deflated: France -0.4\%, Germany 1.4\%, Japan 1.0\%, UK 1.4\%, and US -0.7\%. Own calculations based on data from the \citet{oecd18}.}  House prices have thus grown in Singapore just like in these advanced economies. However, this could only be achieved through cooling measures, which reverted rising into shrinking prices. This will have affected not only average prices, but the whole dwelling value distributions.

\begin{center}
    [\autoref{fig:violin_real_prices} about here.]
\end{center}
\autoref{fig:violin_real_prices} shows violin plots for the dwelling value distribution in the public and the private housing sector. The different deflators simply shift the distributions up- or downwards. In the public sector, the distributions  seem  fairly stable in the inter-quartile range the  central and the median coincide always. However,  the distributions  become compressed and bi-modal  during the cooling rounds. In the private sector, the central value is always below the median. It also appears that the distance of the upper-quartile and the median increases. We assess next how these changes of the dwelling value distributions affect social welfare.  

\subsection{Equivalent housing wealth}

\autoref{fig:equivalent_wealth} gives the estimates the of the equivalent housing wealth ratios $\psi_{0r}$ for each  round $r\geqslant 1$. The cell shading indicates the significance level at which the null hypothesis $\psi_{0r}\geqslant1$ can be rejected. In the few occasions where such a rejection is possible, the significance level is always 1\%. 
\begin{center}
[\autoref{fig:equivalent_wealth} about here.]
\end{center}

The rows with $\nu=0$  corresponds to the case where policy makers do not care about the housing wealth distribution and equivalent equals average housing wealth.
In the public sector, the average housing wealth is smaller in $r=10$ than in the base period for GNI-deflated prices. This is consistent with the values of the GNI-deflated transaction price index in \autoref{fig:average_prices}. Such consistency does not exist for the private sector results when WR- and GNI-deflated prices are used. This is caused by the dwelling composition of transactions and stock being different in the private sector, see the last two columns in \autoref{tab:descriptive_type}. Accounting for the composition of the stock shows that  the average housing wealth in the private sector is lower in $r=5$ and $r=10$ than in the base round. The  rows with $\nu\geqslant1$  correspond to cases where policymakers care about the distribution of housing wealth. In the public sector, the qualitative result remains unchanged and welfare has suffered only in $r=10$ and only so for GNI-deflated prices. In the private sector, however, inequality aversion has an effect on the qualitative results. For instance, if the degree of inequality aversion is high and WR-deflated prices are used, equivalent wealth in $r=10$ is smaller than in the base period. Similar cases occur for GNI-deflated prices.  

The analysis of equivalent wealth indicates that the cooling measures had only a moderate effect on welfare in the public housing sector, which is populated by citizens and permanent residents. Relative to the base period, equivalent wealth is significantly smaller by about 7\% in $r=10$, but  only for GNI-deflated prices. The results are mixed for the private sector, which is populated by foreigners and well-off citizens and permanent residents.  Equivalent wealth is smaller than in the base period  in $r=10$  not only when GNI-deflated prices are used (8\% smaller), but also when the degree of inequality aversion is high and WR-deflated prices are used  (about 3\% smaller).  When GNI-deflated prices are used, equivalent wealth can be smaller than in the base period even for $r\leqslant 9$. 

Given that policymakers in Singapore promote housing wealth as means for citizens to participate in the country's economic growth, the results in \autoref{fig:equivalent_wealth} indicate that unintended effects of cooling measures have been fairly small. However, there is no obvious reason of why policymakers should use the particular function as given in \autoref{eq:functional_form_u}. It seems much more appropriate to suppose that policymakers feel comfortable with $u(p)\in\mathcal{U}_s$, $s=1,2,3$, but not more. As policymakers also desire that welfare improves over time, SD becomes a necessary condition.
 
\subsection{Dominance of housing wealth distributions}

For each $r\geqslant 1$, we apply two different test designs to assess whether and how cooling measures have affected social welfare from  housing wealth. In the first design, we test if $F_r$ SD $F_0$.  If we  reject the null of SD for a particular order $s=1,2,3$, then this implies that social welfare under $F_r$ will not be higher than under $F_0$ for all utility functions $u\in\mathcal{U}_s$. Such an outcome should be of concern to policymakers.
In the second design, we test if  $F_r$ SD $F_k$ for each $r>k\geqslant0$ and thus the wealth distributions from \emph{all} preceding rounds. If we can reject the simultaneous null of SD for a particular $s$, which happens if the hypothesis can be rejected for any of the preceding rounds, then this will be of concern to policymakers.

\begin{center}
[Fig.~\ref{fig:SD_hypothesis1} about here.]
\end{center}
The first column in \autoref{fig:SD_hypothesis1} reports the outcomes from the first test design.\footnote{ \autoref{tab:SD_hypothesis1} in the Appendix gives the $p$-values.} The results for the public sector are qualitatively identical with those from \autoref{fig:equivalent_wealth}. Stochastic dominance for GNI-deflated prices for $r=10$ can be rejected for any order $s=1,2,3$. The necessary conditions for $W_r-W_0>0$ for all $u\in\mathcal{U}_s$ are thus not fulfilled. This is a generalisation  of the results for the five particular utility functions given in \autoref{fig:equivalent_wealth}. In the private sector, the SD tests generalise results from \autoref{fig:equivalent_wealth} in the three cases when GNI-deflated prices are used. The necessary conditions for $W_r>W_0$ for all $u\in\mathcal{U}_s$ can be rejected for $r=4,5,10$.\footnote{The hypothesis of $\psi_{40}\geqslant1$ in \autoref{fig:equivalent_wealth} cannot be rejected for $\nu=0$, which conflicts with the results of the SD tests. The tests for the ratios rely on the delta method, which may cause a loss of power.}  Overall, the SD tests show that welfare from housing wealth in the public sector is not higher in  $r=10$ (the period in which all cooling measures are in place) than in the base period for GNI-deflated prices and all functions in $\mathcal{U}_s$. As policymakers want that citizens and permanent residents participate in the country's economic growth through housing wealth, this outcome might concern them. In the private sector, welfare is not larger than in the base period in $r=4,5,10$.

\begin{center}
[Fig.~\ref{fig:SD_hypothesis2} about here.]
\end{center}
The first panel of \autoref{fig:SD_hypothesis2} shows the results for the second test design. It is obvious that $W_r>W_k$  does not apply for $r>k\geqslant0$ for several rounds, because the necessary conditions  $F_r$ SD $F_k$ can be rejected. Most of the test results in \autoref{fig:SD_hypothesis2} are in line with the behaviour of average prices in \autoref{fig:average_prices}, but the tests provide also additional insights. 
For instance, while GNI-deflated average prices in the public sector reach their peak in 2013Q1 ($r=7$), \autoref{fig:SD_hypothesis2} reveals that on the path to the peak,  $W_r>W_k$ can be rejected for $r=3,4$. While the public sector was only an indirect target in early cooling rounds and average prices where growing, albeit at a low rate, welfare was affected adversely. Similarly, while average CPI-deflated prices peaked at the same time in the public and the private sector, \autoref{fig:SD_hypothesis2} shows that rejections of $W_r>W_k$ are possible in the private sector much earlier  than in the public sector. This differential effect might be expected, given that early cooling measures targeted primarily the private sector. Overall, the first panel in \autoref{fig:SD_hypothesis2} shows that cooling the average house price comes at the cost that welfare from housing wealth is not monotonically increasing. While it might be possible in principle that average prices can be reduced without affecting the level of welfare, this did not have happen in Singapore.

\subsection{Dominance of price trend and unsystematic transaction noise}
Finally, we examine the distribution of level-enhanced residuals to assess the joint effect of cooling measures on the price trend and the unsystematic transaction noise. We assume a notional household who owns the dwelling that underlies the official constant-quality index. The trend measured with the official price index gives the expected value for the price of this standard dwelling. However, there is also transaction noise, which we measure with the residual distribution. The cooling measures can affect this noise. This is therefore the price distribution our notional seller is confronted with. \autoref{fig:violin_level_enhanced_residuals} shows violin plots and the  distributions in the public sector. In the public sector, the distributions around the trend look similar over time except for $r=10$.
\begin{center}
[\autoref{fig:violin_level_enhanced_residuals} about here.]
\end{center}
This cannot be said for the  distributions in the private sector, where the distribution of the unsystematic part changes shape. The notional owner of the standard dwelling implicit in the price trend will evaluate along the expected utility they bring. 

This seller is interested in the expected utility from the price distribution and SD can be tested as before. The right panels in \autoref{fig:SD_hypothesis1} and \autoref{fig:SD_hypothesis2} give the results for the two SD test designs applied to the level-enhanced residuals. There is only one rejection of the necessary conditions for $W_r>W_0$ for all $u\in\mathcal{U}_s$ in \autoref{fig:SD_hypothesis1} , which occurs for GNI-deflated prices in the public sector.  This results is identical to the one we obtain once we test SD for real price levels.
In contrast, the tests  for SD can never be rejected in the private sector. Since the residuals isolate the effects of housing characteristics, the different results between prices and level-enhanced residuals indicate that  adverse changes  in the dwelling distribution are driven by the systematic component of valued characteristics. 

\autoref{fig:SD_hypothesis2} gives in its second panel the results of the second test design for the level-enhanced residuals. There is not much difference between these results and the results for the real prices.

%% file: cooling_conclusion.tex
\section{Conclusion}\label{sec:conclusion}

Every homeowner likes to see home values to rise. However, excessive house price growth and 
any accompanying loan growth can 
``present significant financial stability risks as well as contribute to and worsen an economic downturn'' \citep[p.30]{mreview15_2}. A sudden drop in housing demand or failure of a financial institution overly exposed to the housing sector poses negative  externalities on others and may lead to the collapse of the whole financial system. Macroprudential measures should prevent this by internalising the costs of such externalities. Applied to the housing market, measures consists of limits on loans with respect to size and servicing, minimum cash down-payments, and additional cost on transactions and short holding periods.

Singapore implemented cooling measure soon after its housing market recovered from the financial crisis 07/08. The country faced the additional challenge that the interventions in the housing market should not put too much burden on households owning in the public sector, because the home is their main asset and essential for their retirement. At the same time, the cooling measures should prevent overheating in the private market. We find that the welfare of households in both the public and private sectors has improved despite the introduction of the cooling measures. Our results indicate that the Singaporean agencies involved in this process have been very good at protecting residents housing wealth.

%% file: cooling_appendix.tex
\section{Appendix}

\subsection{Testing for changes of equivalent wealth}\label{app:testing_equivalent_wealth}

We estimate $\psi_{r,0}$ using the transaction price samples for round $r$ of cooling measures and for the base period. For each sample $l\in\{r,0\}$, we compute the utility values $\mathcal{U}_l=\{u(p_{l,1}),\hdots, u(p_{l,N_l})\}$
using \autoref{eq:functional_form_u} and a particular value of the degree of relative inequality aversion $\epsilon\in\{0,1,1.5,2,2.5\}$. The values for $\epsilon$ are standard in the literature, see \citet[pp.127]{lambert01}.
The estimators for social welfare and the variance are
\begin{equation}
 \hat{W}_l=\frac{1}{N_l}\sum_{k=1}^{N_l}\sum_{t=1}^Tu(p_{l,k})\vek{1}(l,k;t)w_l^t
\end{equation}
and, respectively,
\begin{equation}
 \hat{\sigma}^2_{\hat{W}_l}=\frac{1}{N_{l}}\sum_{k=1}^{N_l}\sum_{t=1}^T\{u(p_{l,k})-\hat{W}_l\}^2\vek{1}(l,k;t)w_l^t
\end{equation}
These estimators are similar to those of \citet[p.1055]{cowellflachaire07}, except that we apply re-weighting. The estimator for the wealth ratio is then
\begin{equation}
\hat{\psi}_{\nu,r}=
 \begin{cases}
\left(\dfrac{\hat{W}_r}{\hat{W}_0}\right)^{1/(1-\nu)} & \mbox{for $\nu\geqslant0$ and $\nu\neq 1$}\\[4mm]
\exp\{\hat{W}_r-\hat{W}_0\} & \mbox{for $\nu=1$}
 \end{cases}
\end{equation}
for $r\geqslant1$. 
Using the delta method and exploiting that the samples are independent, the estimator for the variance is
\begin{equation}
 \hat{\sigma}^2_{\hat{\psi}_{\nu,r}}=\begin{cases}
 \left\{\dfrac{1}{(1-\nu)\hat{W}_0}\left(\dfrac{\hat{W}_r}{\hat{W}_0}\right)^{\nu/(1-\nu)}\right\}^2
 \left\{\hat{\sigma}^2_{\hat{W}_r}+\hat{\sigma}^2_{\hat{W}_0}\left(\dfrac{\hat{W}_r}{\hat{W}_0}\right)^2\right\} & \mbox{for $\nu\geqslant0$ and $\nu\neq 1$}\\[0mm]
 \exp\{2(\hat{W}_r-\hat{W}_0)\}\left(\hat{\sigma}^2_{\hat{W}_r}+\hat{\sigma}^2_{\hat{W}_0}\right) & \mbox{for $\nu=1$}
 \end{cases}
\end{equation}

For each bootstrap replication, we draw---with replacement---artificial samples of size $N_l$ from $\mathcal{U}_l$, $l\in\{r,0\}$. The resulting two bootstrap samples are then used to compute $\hat{\psi}_{r,0}^b$ as above. The test statistic in the bootstrap is
\begin{equation}
 \theta_{\nu,r}^b= \frac{(\hat{\psi}^b_{\nu,r}-1)-(\hat{\psi}_{\nu,r}-1)}{\hat{\sigma}^b_{\hat{\psi}_{\nu,r}}}
\end{equation}
It re-centers the estimated bootstrap equivalent wealth ratio with the ratio estimated from the original samples. The null hypothesis is thus imposed in the bootstrap \citep[(16)]{dufouretal18}. After $B$ bootstrap replications, the p-value for the null hypothesis
\begin{equation}
 p_{\nu,r}=\frac{1}{B}\sum_{b=1}^B\vek{1}\left(\hat{\theta}^b_{\nu,r}\leqslant \hat{\theta}_{\nu,r}\right)
\end{equation}
is computed. Alternatively, we can order the test statistics under the null by size and use $c_{\nu,r}(\gamma)=\hat{\theta}^{b}_{\nu,r,[\gamma B]}$ as critical value, where $[\cdot]$ is the integer component.  For instance, for $B=1000$ and $\gamma=0.05$, the critical value at the 5\% level is  $c_{\nu,r}(0.05)=\hat{\theta}^{b}_{\nu,r,50}$. If $\hat{\theta}_{\nu,r}\leqslant c_{\nu,r}(0.05)$, then we reject at the 5\% significance level. Otherwise we do not reject.

\subsection{Testing for stochastic dominance}\label{app:testing_stochastic_dominance}

We estimate $D_{ji}^{(s)}(p)$ in \autoref{eq:estimator_d_hat} with
\begin{equation}
 \hat{D}_{ji}^{(s)}(p)=\hat{D}_{j}^{(s)}(p)-\hat{D}_i^{(s)}(p)
\end{equation}
where $D_l^{(s)}(p)$, $l\in\{i,j\}$ is a functional of $F_l(p)$, see \autoref{subsec:welfare_assessment}. We use the estimator
\begin{equation}\label{eq:estimator_D}
\hat{D}_{l}^{(s)} (p) = \frac{1}{N_l(s-1)!} \sum_{k = 1}^{N_l}\sum_{t=1}^T (p-p_{k,l})^{s-1}\textbf{1}(p_{k,l} \leqslant p)\textbf{1}(k,l;t)w^t_l
\end{equation}
which is similar to the one suggested by \citet{davidsonduclos00}, but applies re-weighting.
To compute the test statistic in \autoref{eq:estimator_d_hat}, we evaluate the estimator in \autoref{eq:estimator_D} at $N_{ji}$ equally spaced points $p$ over the joint range for $j$ and $i$ of observed transaction prices or, respectively, level-enhanced residuals. We conduct inference again with the re-centred bootstrap \citep[Eq.~(11)]{barrettdonald03}. We draw, with replacement, separate bootstrap samples from each of the individual transaction samples  and compute the test statistic
\begin{equation}
\hat{d}_{j}^{(s),b} = \max_{i \neq j} \sup_{p \in \left[0, \bar{p}\right]} {\sqrt{N_{ji}}} \left[\hat{D}_{ji}^{(s),b} (p)-\hat{D}_{ji}^{(s)} (p)\right]
\end{equation}
for each of the 1000 bootstrap replications.

Using the bootstrapped test statistics under the null allows us to calculate p-values
\begin{equation}
  p_j=\frac{1}{B}\sum_{b=1}^B\mathbf{1}\left(\hat{d}^{(s),b}_{j}>\hat{d}^{(s)}_j\right)
\end{equation}
Alternatively, we can order the bootstrapped test statistics ascending by size and use $c_j(\gamma)=\hat{d}^{(s),b}_{j[(1-\gamma)B]}$ as critical value.  For instance, for $B=1000$ and $\gamma=0.05$, the critical value at the 5\% level is $c_j(0.05)=\hat{d}^{(s),b}_{j,950}$. If $\hat{d}^{(s)}_j\geqslant c_j(\gamma)$, the we reject at the $\gamma$ significance level. Otherwise we do not reject. 

\subsection{Results for stochastic dominance tests}

\autoref{tab:SD_hypothesis1} gives the p-values for the tests of stochastic dominance at different degrees for the dwelling prices in levels, deflated with the three different deflators, and for level enhanced residuals.
\begin{center}
    [\autoref{tab:SD_hypothesis1} about here.]
\end{center}
The table complements 
\autoref{fig:SD_hypothesis1} and and \autoref{fig:SD_hypothesis2}, which indicate only the critical levels of significance.

%% file: tables/table_context_descriptives.tex
\begin{table}[ph!]
\centering
\caption{{\bf Ownership rates, assets and liabilities, Singapore 2009-2016.} Figures are in percent. Panel A relates the number of owner-occupying resident households to the total number of households living in particular property type. Panel B relates types of assets to asset total, Panel C does the same for liabilities. Mortgage ratio is the amount of mortgage liabilities to the value of residential property assets. Figures in Panel A are based on authors' calculations using information from  \citet[Tables 4.2 and 4.4]{statyearbook16} and \citet[Tables 3.2 and 3.4]{statyearbook17}. Figures in Panels B and C are based on authors' calculations using
year-end values of household sector balance sheet M700981.}
\label{tab:descriptives_context}
\begin{adjustbox}{max width = \textwidth}
\begin{tabular}{lrrrrrrrr}
\toprule
 & \multicolumn{1}{c}{\bf 2009} & \multicolumn{1}{c}{\bf 2010} & \multicolumn{1}{c}{\bf 2011} &	\multicolumn{1}{c}{\bf 2012} &	\multicolumn{1}{c}{\bf 2013} &	\multicolumn{1}{c}{\bf 2014} &	\multicolumn{1}{c}{\bf 2015} & 	\multicolumn{1}{c}{\bf 2016} \\\cline{1-9}
  & \multicolumn{8}{c}{Panel A: Ownership rates}\\\cline{2-9}							
  All	                                 &88.8	&87.2	&88.6	&90.1	&90.5	&90.3	&90.8	&90.9 \\
  \quad HDB all flat types	             & 90.4	&88.8	&90.1	&91.7	&91.8	&91.6	&92.0	&92.2 \\
  \quad HDB larger than 2 rooms          & 94.5	&92.9	&94.3	&95.9	&96.3	&96.4	&96.4	&96.7 \\
  Private	                             & 80.9	&79.7	&81.4	&83.0	&84.5	&85.0	&85.8	&85.9 \\\cline{2-9}
  &	\multicolumn{8}{c}{Panel B: Assets}	\\\cline{2-9}
  Financial assets	                     &53.0	&50.9	&49.7	&50.5	&51.7	&53.6	&54.2	&55.1 \\
  Residential property 	                 &47.0	&49.1	&50.3	&49.5	&48.3	&46.4	&45.8	&44.9 \\
  \quad Public housing	                 &24.9	&24.9	&25.8	&25.3	&24.1	&22.4	&22.0	&21.5 \\
  \quad Private housing	                 &22.1	&24.2	&24.5	&24.2	&24.2	&24.1	&23.7	&23.3 \\\cline{2-9}
  &	\multicolumn{8}{c}{Panel C: Liabilities}	\\\cline{2-9}							
  Mortgage loans	                     &73.7	&74.8	&73.8	&73.4	&73.3	&73.6	&74.5	&75.4 \\
  \quad Financial institutions	         &49.6	&54.5	&56.5	&58.6	&60.2	&61.0	&62.0	&62.8 \\
  \quad HDB	                             &24.1	&20.3	&17.3	&14.8	&13.1	&12.6	&12.6	&12.6 \\
  Personal loans                         &26.3	&25.2	&26.2	&26.6	&26.7	&26.4	&25.5	&24.6 \\\cline{2-9}
Mortgage ratio	                         &24.3	&23.0	&23.0	&23.7	&24.8	&26.5	&27.0	&27.0 \\
Liabilities to assets	                 &15.5	&15.1	&15.7	&16.0	&16.3	&16.7	&16.6	&16.1 \\
\bottomrule
\end{tabular}
\end{adjustbox}
\end{table} 

%% file: tables/table_n_obs.tex
\begin{table}[ph!]
\centering
\caption{{\bf Cooling measure rounds and dwelling transactions,  2009M2-2016M6.} Shows the dates when the different rounds of cooling measures were launched. Days count how long it took until a new round was launched. Round 0 is the base period and covers the time from the start of our transaction data set to the day before Round 1 of cooling measures was launched. Round 10 ends with the end of our transaction data sample. Rounds 11 and 12 introduced only minor changes to existing measures and we subsume them under Round 10. The last three columns show the number of observations of our transaction data set in total and split by the public HDB resale and private sector.}
\label{tab:n_obs}
\begin{adjustbox}{max width = \textwidth}
\begin{tabular}{@{}lrrrrrr@{}}
\toprule
{\bf Round} & \multicolumn{1}{c}{\bf Launch} &  \multicolumn{1}{c}{\bf Days} & \multicolumn{1}{c}{\bf Public} & \multicolumn{1}{c}{\bf Private} & \multicolumn{1}{c}{\bf Total} \\ \midrule
0 & 1/2/09 & 225 & 17,118 & 22,258 & 39,376 \\
1 & 14/9/09 & 159 & 9,982 & 13,530 & 23,512 \\
2 & 20/2/10 & 191 & 12,595 & 20,185 & 32,780 \\
3 & 30/8/10 & 137 & 6,822 & 12,631 & 19,453 \\
4 & 14/1/11 & 328 & 20,547 & 28,387 & 48,934 \\
5 & 8/12/11 & 303 & 20,511 & 28,316 & 48,827 \\
6 & 6/10/12 & 98 & 5,735 & 9,119 & 14,854 \\
7 & 12/1/13 & 168 & 8,173 & 12,936 & 21,109 \\
8 & 29/6/13 & 59 & 2,666 & 2,832 & 5,498 \\
9 & 27/8/13 & 104 & 4,268 & 4,766 & 9,034 \\
10 & 9/12/13 & 935 & 46,630 & 34,201 & 80,831 \\ \midrule
\multicolumn{1}{r}{\bf Total} & & 2,707 & 155,047 & 189,161 & 344,208 \\ \bottomrule
\end{tabular}
\end{adjustbox}
\end{table} 

%% file: tables/descriptive_type.tex
\begin{table}[p]
\centering
\caption{{\bf Descriptive statistics for transaction data set, 2009M2-2016M6.} Reports average of a variable and standard deviation in bracket for continuous variables.  Public sector transactions are always resales and leaseholds. Resale and leasehold are the base categories for private sector transaction. New indicates the sale of a new dwelling from a developer and subsale refers to the consecutive sale of an uncompleted dwelling. The transaction coverage ratio relates the number of transactions of a dwelling type  to the total  transactions. The stock coverage ratio relates the number of a dwelling type in the total stock. Stock refers to all dwellings, regardless of whether they are owner-occupied, rented-out, or vacant. }
\label{tab:descriptive_type}
\begin{adjustbox}{max width = \linewidth}
\begin{threeparttable}
\begin{tabular}{@{}lrrrrrrrrrrrr@{}}
\toprule
\multicolumn{1}{c}{} & \multicolumn{5}{c}{\bf Public sector} & \multicolumn{1}{c}{} & \multicolumn{6}{c}{\bf Private sector} \\ \cmidrule(lr){2-6} \cmidrule(l){8-13}
&\multicolumn{1}{c}{Overall} &\multicolumn{1}{c}{3 Rooms} & \multicolumn{1}{c}{4 Rooms} & \multicolumn{1}{c}{5 Rooms} & \multicolumn{1}{c}{Executive} & &\multicolumn{1}{c}{Overall} &\multicolumn{2}{c}{Multi-family structure} & \multicolumn{3}{c}{Single-family structure}\\ \cmidrule(lr){2-6}\cmidrule(l){8-13}
\multicolumn{1}{c}{} & \multicolumn{1}{c}{} & \multicolumn{1}{c}{} & \multicolumn{1}{c}{} & \multicolumn{1}{c}{} & \multicolumn{1}{c}{} & \multicolumn{1}{c}{} & \multicolumn{1}{c}{} &\multicolumn{1}{c}{Apartment} & \multicolumn{1}{c}{Condominium} & \multicolumn{1}{c}{Detached} & \multicolumn{1}{c}{Semi-Detached} & \multicolumn{1}{c}{Terrace} \\ \cmidrule(r){1-6} \cmidrule(l){8-13}
Price (S\$) & 424,103.97 & 321,669.04 & 420,164.95 & 497,022.75 & 583,581.20 &  & 1,499,699.01 & 1,198,871.02 & 1,403,770.99 & 7,288,856.28 & 3,336,735.03 & 2,125,399.65 \\
 & (117,679.41) & (57,673.16) & (89,133.78) & (106,172.80) & (97,086.34) &  & (1,312,283.38) & (899,234.09) & (1,063,922.55) & (4,376,257.17) & (1,461,388.79) & (866,408.79) \\
Area (m$^2$) & 97.39 & 68.08 & 96.22 & 118.31 & 143.04 &  & 123.56 & 88.02 & 113.14 & 660.06 & 336.94 & 215.57 \\
 & (23.89) & 6.36 & (8.11) & (7.32) & (8.58) &  & (92.74) & (46.09) & (50.20) & (258.80) & (100.38) & (78.57) \\
Storey & 7.68 & 6.84 & 7.70 & 8.76 & 7.17 &  & 8.15 & 8.96 & 9.04 &  &  &  \\
 & (4.75) & (3.90) & (4.87) & (5.36) & (4.20) &  & (7.49) & (7.73) & (7.20) &  &  &  \\
Distance to City Hall (km) & 11.88 & 9.82 & 12.33 & 13.01 & 13.47 &  & 8.24 & 6.07 & 9.10 & 7.83 & 9.06 & 9.45 \\
 & (4.40) & (4.35) & (4.27) & (4.11) & (3.34) &  & (4.33) & (3.97) & (4.28) & (3.21) & (3.10) & (3.43) \\
New &  &  &  &  &  &  & .50 & .55 & .53 & .09 & .13 & .14 \\
Subsale &  &  &  &  &  &  & .08 & .09 & .08 & .02 & .01 & .02 \\
Freehold &  &  &  &  &  &  & .45 & .63 & .32 & .88 & .89 & .76 \\
Observations & \multicolumn{1}{r}{155,047} & \multicolumn{1}{r}{43,707} & \multicolumn{1}{r}{61,206} & \multicolumn{1}{r}{37,859} & \multicolumn{1}{r}{12,275} & \multicolumn{1}{r}{} & \multicolumn{1}{r}{189,161} & \multicolumn{1}{r}{53,885} & \multicolumn{1}{r}{117,189} & \multicolumn{1}{r}{1,968} & \multicolumn{1}{r}{4,931} & \multicolumn{1}{r}{11,188} \\
Transaction coverage               &           & .28      & .40      & .25       & .08       &  &             & .28       & .62         & .01         & .03           & .06       \\
Stock coverage                &           & .26      & .42      & .25       & .07       &  &             & .26       & .47         & .04         & .08           & .15
\\ \bottomrule
\end{tabular}
\end{threeparttable}
\end{adjustbox}
\end{table} 

%% file: tables/SD_hypothesis1.tex
\begin{table}[htp]
\centering
\caption{{\bf Stochastic dominance tests.} Shows in the first panel the p-values for the null hypotheses that the dwelling value distribution from round $r\geqslant1$ dominates the base round distribution and in the second round the p-values for the hypothesis that round $r\geqslant1$ dominates all preceding rounds.}
\label{tab:SD_hypothesis1}
\begin{adjustbox}{max width = 0.85\linewidth}
\begin{tabular}{@{}llSSSSSSSSSSSSSSSSSSSSS@{}}
\toprule
\multicolumn{1}{c}{} & \multicolumn{1}{c}{} & \multicolumn{10}{c}{Real prices} & \multicolumn{1}{c}{} & \multicolumn{10}{c}{Level-enhanced residuals} \\ \cmidrule(lr){3-12} \cmidrule(lr){14-23}
\multicolumn{1}{c}{} & \multicolumn{1}{c}{} & \multicolumn{1}{c}{1} & \multicolumn{1}{c}{2} & \multicolumn{1}{c}{3} & \multicolumn{1}{c}{4} & \multicolumn{1}{c}{5} & \multicolumn{1}{c}{6} & \multicolumn{1}{c}{7} & \multicolumn{1}{c}{8} & \multicolumn{1}{c}{9} & \multicolumn{1}{c}{10} & \multicolumn{1}{c}{} & \multicolumn{1}{c}{1} & \multicolumn{1}{c}{2} & \multicolumn{1}{c}{3} & \multicolumn{1}{c}{4} & \multicolumn{1}{c}{5} & \multicolumn{1}{c}{6} & \multicolumn{1}{c}{7} & \multicolumn{1}{c}{8} & \multicolumn{1}{c}{9} & \multicolumn{1}{c}{10} \\\midrule\multicolumn{23}{c}{Stochastic dominance of round $r$ over the base round 0}\\\midrule
\multicolumn{2}{l}{Public, CPI}  &  &  &  &  &  &  &  &  &  &  &  &  &  &  &  &  &  &  &  &  &  \\
 & FSD & 1.000 & 1.000 & 1.000 & 1.000 & 1.000 & 1.000 & 1.000 & 1.000 & 1.000 & 1.000 &  & 1.000 & 1.000 & 1.000 & 1.000 & 1.000 & 1.000 & 1.000 & 1.000 & 1.000 & 1.000 \\
 & SSD & 1.000 & 1.000 & 1.000 & 1.000 & 1.000 & 1.000 & 1.000 & 1.000 & 1.000 & 1.000 &  & 1.000 & 1.000 & 1.000 & 1.000 & 1.000 & 1.000 & 1.000 & 1.000 & 1.000 & 1.000 \\
 & TSD & 1.000 & 1.000 & 1.000 & 1.000 & 1.000 & 1.000 & 1.000 & 1.000 & 1.000 & 1.000 &  & .988 & .991 & .987 & 1.000 & .993 & .996 & .995 & .985 & .988 & 1.000 \\
\multicolumn{2}{l}{Public, Wage}  &  &  &  &  &  &  &  &  &  &  &  &  &  &  &  &  &  &  &  &  &  \\
 & FSD & 1.000 & 1.000 & 1.000 & 1.000 & 1.000 & 1.000 & 1.000 & 1.000 & 1.000 & .855 &  & 1.000 & 1.000 & 1.000 & 1.000 & 1.000 & 1.000 & 1.000 & 1.000 & 1.000 & 1.000 \\
 & SSD & 1.000 & 1.000 & 1.000 & 1.000 & 1.000 & 1.000 & 1.000 & 1.000 & 1.000 & 1.000 &  & 1.000 & 1.000 & 1.000 & 1.000 & 1.000 & 1.000 & 1.000 & 1.000 & 1.000 & 1.000 \\
 & TSD & 1.000 & 1.000 & 1.000 & 1.000 & 1.000 & 1.000 & 1.000 & 1.000 & 1.000 & 1.000 &  & .992 & .993 & .993 & .997 & .993 & .989 & .988 & .984 & .986 & 1.000 \\
\multicolumn{2}{l}{Public, GNI}  &  &  &  &  &  &  &  &  &  &  &  &  &  &  &  &  &  &  &  &  &  \\
 & FSD & 1.000 & 1.000 & .996 & .753 & 1.000 & 1.000 & 1.000 & 1.000 & 1.000 & .000 &  & 1.000 & 1.000 & 1.000 & 1.000 & 1.000 & 1.000 & 1.000 & 1.000 & 1.000 & .000 \\
 & SSD & 1.000 & 1.000 & 1.000 & 1.000 & 1.000 & 1.000 & 1.000 & 1.000 & 1.000 & .000 &  & 1.000 & 1.000 & 1.000 & 1.000 & 1.000 & 1.000 & 1.000 & 1.000 & 1.000 & .000 \\
 & TSD & 1.000 & 1.000 & 1.000 & 1.000 & 1.000 & 1.000 & 1.000 & 1.000 & 1.000 & .000 &  & .995 & .995 & .992 & .993 & .990 & .989 & .994 & .989 & .990 & .000 \\
\multicolumn{2}{l}{Private, CPI}  &  &  &  &  &  &  &  &  &  &  &  &  &  &  &  &  &  &  &  &  &  \\
 & FSD & 1.000 & 1.000 & 1.000 & 1.000 & .809 & 1.000 & 1.000 & 1.000 & 1.000 & 1.000 &  & 1.000 & 1.000 & 1.000 & 1.000 & 1.000 & 1.000 & 1.000 & 1.000 & 1.000 & 1.000 \\
 & SSD & 1.000 & 1.000 & 1.000 & 1.000 & .776 & 1.000 & 1.000 & 1.000 & .989 & 1.000 &  & 1.000 & 1.000 & 1.000 & 1.000 & 1.000 & 1.000 & 1.000 & 1.000 & 1.000 & 1.000 \\
 & TSD & 1.000 & 1.000 & 1.000 & 1.000 & .709 & 1.000 & 1.000 & 1.000 & .982 & 1.000 &  & 1.000 & 1.000 & 1.000 & 1.000 & 1.000 & 1.000 & 1.000 & 1.000 & 1.000 & 1.000 \\
\multicolumn{2}{l}{Private, Wage}  &  &  &  &  &  &  &  &  &  &  &  &  &  &  &  &  &  &  &  &  &  \\
 & FSD & 1.000 & 1.000 & 1.000 & .258 & .000 & 1.000 & .819 & 1.000 & .667 & .000 &  & 1.000 & 1.000 & 1.000 & 1.000 & 1.000 & 1.000 & 1.000 & 1.000 & 1.000 & 1.000 \\
 & SSD & 1.000 & 1.000 & 1.000 & .689 & .270 & 1.000 & .729 & 1.000 & .701 & .073 &  & 1.000 & 1.000 & 1.000 & 1.000 & 1.000 & 1.000 & 1.000 & 1.000 & 1.000 & 1.000 \\
 & TSD & 1.000 & 1.000 & 1.000 & .676 & .448 & 1.000 & .704 & 1.000 & .675 & .149 &  & 1.000 & 1.000 & 1.000 & 1.000 & 1.000 & 1.000 & 1.000 & 1.000 & 1.000 & 1.000 \\
\multicolumn{2}{l}{Private, GNI}  &  &  &  &  &  &  &  &  &  &  &  &  &  &  &  &  &  &  &  &  &  \\
 & FSD & 1.000 & .975 & .108 & .000 & .000 & .000 & .000 & .064 & .000 & .000 &  & 1.000 & 1.000 & 1.000 & 1.000 & 1.000 & 1.000 & 1.000 & 1.000 & 1.000 & 1.000 \\
 & SSD & 1.000 & .834 & .478 & .019 & .000 & .293 & .069 & .400 & .131 & .000 &  & 1.000 & 1.000 & 1.000 & 1.000 & 1.000 & 1.000 & 1.000 & 1.000 & 1.000 & 1.000 \\
 & TSD & 1.000 & .768 & .534 & .040 & .000 & .401 & .172 & .460 & .189 & .000 &  & 1.000 & 1.000 & 1.000 & 1.000 & 1.000 & 1.000 & 1.000 & 1.000 & 1.000 & .998\\
 \midrule\multicolumn{23}{c}{Stochastic dominance of round $r$  over all preceding rounds}\\\midrule
 \multicolumn{2}{l}{Public, CPI}  &  &  &  &  &  &  &  &  &  &  &  &  &  &  &  &  &  &  &  &  &  \\
 & FSD & 1.000 & 1.000 & 1.000 & 1.000 & 1.000 & 1.000 & .992 & .112 & .000 & .000 &  & 1.000 & 1.000 & 1.000 & 1.000 & 1.000 & 1.000 & 1.000 & .000 & .000 & .000 \\
 & SSD & 1.000 & 1.000 & 1.000 & 1.000 & 1.000 & 1.000 & 1.000 & .140 & .000 & .000 &  & 1.000 & 1.000 & 1.000 & 1.000 & 1.000 & 1.000 & 1.000 & .000 & .000 & .000 \\
 & TSD & 1.000 & 1.000 & 1.000 & 1.000 & 1.000 & 1.000 & 1.000 & .070 & .000 & .000 &  & .988 & 1.000 & 1.000 & 1.000 & 1.000 & 1.000 & 1.000 & .000 & .000 & .000 \\
\multicolumn{2}{l}{Public, Wage}  &  &  &  &  &  &  &  &  &  &  &  &  &  &  &  &  &  &  &  &  &  \\
 & FSD & 1.000 & 1.000 & 1.000 & 1.000 & 1.000 & 1.000 & .010 & .000 & .000 & .000 &  & 1.000 & 1.000 & 1.000 & 1.000 & 1.000 & 1.000 & .000 & .000 & .000 & .000 \\
 & SSD & 1.000 & 1.000 & 1.000 & 1.000 & 1.000 & 1.000 & .548 & .027 & .000 & .000 &  & 1.000 & 1.000 & 1.000 & 1.000 & 1.000 & 1.000 & .000 & .000 & .000 & .000 \\
 & TSD & 1.000 & 1.000 & 1.000 & 1.000 & 1.000 & 1.000 & .597 & .001 & .000 & .000 &  & .992 & 1.000 & 1.000 & 1.000 & 1.000 & 1.000 & .000 & .000 & .000 & .000 \\
\multicolumn{2}{l}{Public, GNI}  &  &  &  &  &  &  &  &  &  &  &  &  &  &  &  &  &  &  &  &  &  \\
 & FSD & 1.000 & .289 & .000 & .000 & .013 & 1.000 & .836 & .001 & .000 & .000 &  & 1.000 & .003 & .000 & .000 & 1.000 & 1.000 & 1.000 & .000 & .000 & .000 \\
 & SSD & 1.000 & 1.000 & .000 & .000 & .730 & 1.000 & .984 & .003 & .000 & .000 &  & 1.000 & .253 & .000 & .000 & 1.000 & 1.000 & 1.000 & .000 & .000 & .000 \\
 & TSD & 1.000 & 1.000 & .000 & .000 & 1.000 & 1.000 & .989 & .001 & .000 & .000 &  & .995 & .964 & .000 & .000 & 1.000 & 1.000 & 1.000 & .000 & .000 & .000 \\
\multicolumn{2}{l}{Private, CPI}  &  &  &  &  &  &  &  &  &  &  &  &  &  &  &  &  &  &  &  &  &  \\
 & FSD & 1.000 & .873 & .915 & .021 & .000 & 1.000 & .073 & .382 & .039 & .000 &  & 1.000 & 1.000 & 1.000 & 1.000 & .000 & .220 & .001 & .004 & .000 & .000 \\
 & SSD & 1.000 & .917 & .999 & .544 & .001 & 1.000 & .234 & .511 & .020 & .000 &  & 1.000 & 1.000 & 1.000 & 1.000 & .000 & 1.000 & 1.000 & 1.000 & .000 & .000 \\
 & TSD & 1.000 & .902 & .999 & .700 & .001 & 1.000 & .020 & .706 & .004 & .000 &  & 1.000 & 1.000 & 1.000 & 1.000 & .000 & 1.000 & 1.000 & 1.000 & .015 & .000 \\
\multicolumn{2}{l}{Private, Wage}  &  &  &  &  &  &  &  &  &  &  &  &  &  &  &  &  &  &  &  &  &  \\
 & FSD & 1.000 & .322 & .042 & .000 & .000 & .050 & .000 & .000 & .001 & .000 &  & 1.000 & 1.000 & 1.000 & 1.000 & .000 & .000 & .000 & .000 & .000 & .000 \\
 & SSD & 1.000 & .799 & .217 & .002 & .000 & .787 & .047 & .122 & .001 & .000 &  & 1.000 & 1.000 & 1.000 & 1.000 & .022 & 1.000 & .000 & .000 & .000 & .000 \\
 & TSD & 1.000 & .818 & .219 & .000 & .000 & .919 & .000 & .165 & .000 & .000 &  & 1.000 & 1.000 & 1.000 & 1.000 & .001 & 1.000 & .000 & .000 & .000 & .000 \\
\multicolumn{2}{l}{Private, GNI}  &  &  &  &  &  &  &  &  &  &  &  &  &  &  &  &  &  &  &  &  &  \\
 & FSD & 1.000 & .002 & .000 & .000 & .000 & .000 & .000 & .000 & .000 & .000 &  & 1.000 & 1.000 & .000 & .000 & .000 & .000 & .000 & .000 & .000 & .000 \\
 & SSD & 1.000 & .007 & .000 & .000 & .000 & .000 & .000 & .002 & .000 & .000 &  & 1.000 & 1.000 & .000 & .000 & .000 & .000 & .000 & .000 & .000 & .000 \\
 & TSD & 1.000 & .000 & .000 & .000 & .000 & .000 & .000 & .000 & .000 & .000 &  & 1.000 & 1.000 & .000 & .000 & .000 & .000 & .000 & .000 & .000 & .000\\ \bottomrule
\end{tabular}
\end{adjustbox}
\end{table}

%% file: singapore.bib
@Article{atkinson08,
  Title                    = {More on the measurement of inequality},
  Author                   = {Anthony B. Atkinson},
  Journal                  = {Journal of Economic Inequality},
  Year                     = {2008},
  Note                     = {Reprint of working paper from 1973},
  Pages                    = {277-283},
  Volume                   = {6}
}

@Article{atkinson70,
  Title                    = {On the Measurement of Inequality},
  Author                   = {Anthony B. Atkinson},
  Journal                  = {Journal of Economic Theory},
  Year                     = {1970},
  Pages                    = {244-263},
  Volume                   = {2}
}

@Manual{boe11,
  Title                    = {Instruments of Macroprudential Policy. A Discussion Paper},

  Address                  = {London},
  Author                   = {{Bank of England}},
  Month                    = {December},
  Year                     = {2011}
}

@Article{barrettdonald03,
  Title                    = {Consistent tests for stochastic dominance},
  Author                   = {Garry F. Barrett and Stephen G. Donald},
  Journal                  = {Econometrica},
  Year                     = {2003},
  Pages                    = {71-104},
  Volume                   = {71}
}

@Article{bawa75,
  Title                    = {Optimal rules for ordering uncertain prospects},
  Author                   = {Vijay S. Bawa},
  Journal                  = {Journal of Financial Economics},
  Year                     = {1975},
  Pages                    = {95-121},
  Volume                   = {2}
}

@Article{brunnermeier09,
  Title                    = {Deciphering the Liquidity and Credit Crunch 2007-2008},
  Author                   = {Markus K. Brunnermeier},
  Journal                  = {Journal of Economic Perspectives},
  Year                     = {2009},
  Number                   = {1},
  Pages                    = {77-100},
  Volume                   = {23}
}

@Article{ceruttietal17,
  Title                    = {The use and effectiveness of macroprudential policies: New evidence},
  Author                   = {Eugenio Cerutti and Stijn Claessens and Luc Laeven},
  Journal                  = {Journal of Financial Stability},
  Year                     = {2017},
  Pages                    = {203-224},
  Volume                   = {28}
}

@Article{cowellflachaire07,
  Title                    = {Income distribution and inequality measurement: The problem of extreme values},
  Author                   = {Frank A. Cowell and Emmanuel Flachaire},
  Journal                  = {Journal of Econometrics},
  Year                     = {2007},
  Pages                    = {1044-1072},
  Volume                   = {141}
}

@TechReport{darbarwu15,
  Title                    = {Experiences with macroprudential policy---five case studies},
  Author                   = {Salim M. Darbar and Xiaoyong Wu},
  Institution              = {International Monetary Fund},
  Year                     = {2015},

  Address                  = {Washington},
  Month                    = {June},
  Number                   = {WP/15/123},
  Type                     = {{IMF} Working Paper}
}

@Article{davidsonduclos00,
  Title                    = {Statistical Inference for Stochastic Dominance and for the Measurement of Poverty and Inequality},
  Author                   = {Russell Davidson and Jean-Yves Duclos},
  Journal                  = {Econometrica},
  Year                     = {2000},
  Pages                    = {1435-1464},
  Volume                   = {68}
}

@Manual{statyearbook17,
  Title                    = {Yearbook of Statistics {S}ingapore},
  Author                   = {{DOS}},
  Organization             = {{Department of Statistics Singapore}},
  Year                     = {2017}
}

@Manual{statyearbook16,
  Title                    = {Yearbook of Statistics {S}ingapore},
  Author                   = {{{DOS}}},
  Organization             = {{Department of Statistics Singapore}},
  Year                     = {2016}
}

@Article{dufouretal18,
  Title                    = {Permutation tests for comparing inequality measures},
  Author                   = {Jean-Marie Dufour and Emmanuel Flachaire and Lynda Khalaf},
  Journal                  = {Journal of Business \& Economic Statistics},
  Year                     = {2018},
  Note                     = {forthcoming}
}

@Article{galatimoessner13,
  Title                    = {Macroprudential Policy --- A Literature Review},
  Author                   = {Gabriele Galati and Richhild Moessner},
  Journal                  = {Journal of Economic Surveys},
  Year                     = {2013},
  Pages                    = {846-878},
  Volume                   = {27}
}

@Article{hansonetal11,
  Title                    = {A macroprudential approach to financial regulation},
  Author                   = {Samuel G. Hanson and Anil K. Kashyap and Jeremy C. Stein},
  Journal                  = {Journal of Economic Perspectives},
  Year                     = {2011},
  Number                   = {1},
  Pages                    = {3-28},
  Volume                   = {25}
}

@TechReport{klecanetal91,
  Title                    = {A robust test for stochastic dominance},
  Author                   = {Lindsay Klecan and Raymond Mc{F}adden and Daniel Mc{F}adden},
  Institution              = {MIT},
  Year                     = {1991},

  Address                  = {Cambridge MA},
  Month                    = {January},
  Type                     = {Working Paper}
}

@Article{kuttnershim16,
  Title                    = {Can non-interest rate policies stabilize housing markets? {E}vidence from a panel of 57 economies},
  Author                   = {Kenneth N. Kuttner and Ilhyock Shim},
  Journal                  = {Journal of Financial Stability},
  Year                     = {2016},
  Pages                    = {31-44},
  Volume                   = {26},

  ISSN                     = {1572-3089},
  Shorttitle               = {Can non-interest rate policies stabilize housing markets?},
  Urldate                  = {2017-02-07TZ}
}

@Book{lambert01,
  Title                    = {The distribution and redistribution of income},
  Author                   = {Peter J. Lambert},
  Publisher                = {Manchester University Press},
  Year                     = {2001},

  Address                  = {Manchester and New York},
  Edition                  = {third}
}

@Article{lintonetal05,
  Title                    = {Consistent Testing for Stochastic Dominance under General Sampling Schemes},
  Author                   = {Oliver Linton and Esfandiar Maasoumi and Yoon-Jae Whang},
  Journal                  = {The Review of Economic Studies},
  Year                     = {2005},
  Pages                    = {735-765},
  Volume                   = {72},

  Language                 = {English}
}

@Article{mreview15_2,
  Title                    = {Housing and Business Cycles in {S}ingapore ({B}ox {A})},
  Author                   = {{MAS}},
  Journal                  = {Monetary Authority Singapore: Macroeconomic Review},
  Year                     = {2015},

  Month                    = {October},
  Number                   = {2},
  Pages                    = {30-33},
  Volume                   = {14}
}

@Manual{masfsr_11,
  Title                    = {Financial Stability Review},
  Author                   = {{MAS}},
  Month                    = {November},
  Note                     = {November 2011},
  Organization             = {{Monetary Authority of Singapore}},
  Year                     = {2011}
}

@Manual{MAS_TER,
  Title                    = {Tenets of Effective Regulation},
  Author                   = {{MAS}},
  Note                     = {June 2010 (revised in {S}eptember 2015)},
  Organization             = {{Monetary Authority of Singapore}},
  Year                     = {2010}
}

@Electronic{oecd18,
  Title                    = {Housing prices (indicator)},
  Author                   = {{OECD}},
  Note                     = {Accessed on 17 {O}ctober 2018},
  Url                      = {https://doi.org/10.1787/63008438-en},
  Year                     = {2018}
}

@InCollection{phang07,
  Title                    = {The {S}ingapore Model of Housing and the Welfare State},
  Author                   = {Sock-Yong Phang},
  Booktitle                = {Housing and the new welfare state: {P}erspectives from {E}ast {A}sia and {E}urope},
  Publisher                = {Routledge},
  Year                     = {2007},

  Address                  = {Abingdon},
  Chapter                  = {2},
  Editor                   = {Richard Groves and Alan Murie and Christopher Watson},
  Pages                    = {15-44},
  Series                   = {Social Policy in Modern {A}sia}
}

@Book{ruppertetal03,
  Title                    = {Semiparametric Regression},
  Author                   = {David Ruppert and M. P. Wand and R. J. Carroll},
  Publisher                = {Cambridge University Press},
  Year                     = {2003},

  Address                  = {Cambridge},
  Series                   = {Cambridge Series in Statistical and Probabilistic Mathematics}
}

@Article{shorrocksforster87,
  Title                    = {Transfer sensitive inequality measures},
  Author                   = {Anthony F. Shorrocks and James E. Forster},
  Journal                  = {Review of Economic Studies},
  Year                     = {1987},
  Pages                    = {485-497},
  Volume                   = {54}
}

@Manual{wandetal05,
  Title                    = {Semi{P}ar 1.0 {R} package},
  Author                   = {M. P. Wand and B. A. Coull and J. L. French and B. Ganguli and E. E. Kammann and J. Staudenmayer and A. Zanobetti},
  Year                     = {2005},
  Url                      = {http://cran.r-project.org}
}

@Article{zhangzoli16,
  Title                    = {Leaning against the wind: Macroprudential policy in {Asia}},
  Author                   = {Longmei Zhang and Edda Zoli},
  Journal                  = {Journal of Asian Economics},
  Year                     = {2016},
  Pages                    = {33-52},
  Volume                   = {42},
  ISSN                     = {1049-0078}
}

@Book{guisoetal02,
  Title                    = {Household Portfolios},
  Editor                   = {Luigi Guiso and Michael Haliassos and Tullio Jappelli},
  Publisher                = {MIT Press},
  Year                     = {2002},

  Address                  = {Cambridge MA}
}
